%% file: frame.tex
\documentclass[sigconf]{acmart}

\usepackage{hyperref}
\usepackage{hyperxmp}
\usepackage{threeparttable}
\usepackage{multirow}
\usepackage{makecell}
\usepackage{colortbl}

\usepackage{amsthm,amsmath,amssymb}
\usepackage{mathrsfs}
\usepackage{subcaption}
\usepackage{graphicx}
\usepackage{enumitem}
\usepackage{bbding}
\usepackage{twemojis}
\usepackage{tablefootnote}
\usepackage{siunitx}
\AtBeginDocument{%
  }

\copyrightyear{2025}
\acmYear{2025}
\setcopyright{cc}
\setcctype{by}
\acmConference[SIGIR '25]{Proceedings of the 48th International ACM SIGIR Conference on Research and Development in Information Retrieval}{July 13--18, 2025}{Padua, Italy}
\acmBooktitle{Proceedings of the 48th International ACM SIGIR Conference on Research and Development in Information Retrieval (SIGIR '25), July 13--18, 2025, Padua, Italy}\acmDOI{10.1145/3726302.3730083}
\acmISBN{979-8-4007-1592-1/2025/07}
\begin{document}

\input{Header/1.title_author}
\input{Header/2.abstract}

\input{Header/3.ccs_keywords}

% \begin{teaserfigure}
%   \includegraphics[width=\textwidth]{sampleteaser}
%   \caption{Seattle Mariners at Spring Training, 2010.}
%   \Description{Enjoying the baseball game from the third-base
%   seats. Ichiro Suzuki preparing to bat.}
%   \label{fig:teaser}
% \end{teaserfigure}

% \received{20 February 2007}
% \received[revised]{12 March 2009}
% \received[accepted]{5 June 2009}

%%
%% This command processes the author and affiliation and title
%% information and builds the first part of the formatted document.
\maketitle
\input{Main/1.Introduction}

\input{Main/2.Related_work}
\input{Main/3.Method}
\input{Main/4.Task1}
\input{Main/5.Task2}
\input{Main/6.Case}
\input{Main/7.Conclusion}

%%
%% The acknowledgments section is defined using the "acks" environment
%% (and NOT an unnumbered section). This ensures the proper
%% identification of the section in the article metadata, and the
%% consistent spelling of the heading.
\begin{acks}
This work is supported by the Natural Science Foundation of China (Grant No. U21B2026, 62372260) and Quan Cheng Laboratory (Grant No. QCLZD202301). 
\end{acks}

%%
%% The next two lines define the bibliography style to be used, and
%% the bibliography file.
\bibliographystyle{ACM-Reference-Format}
\bibliography{reference}

%%
%% If your work has an appendix, this is the place to put it.

\end{document}

%% file: Header/1.title_author.tex
%% The "title" command has an optional parameter,
%% allowing the author to define a "short title" to be used in page headers.

\title{Short Video Segment-level User Dynamic Interests Modeling in Personalized Recommendation}

% Segment-level User Dynamic Interests Modeling in Video Recommendation
% Short Video Segment-level User Dynamic Interests Modeling and Its application for Recommendation
% Short Video Segment-level User Dynamic Interests Modeling and Recommendation

% 不用强调short video？
% clip-based： 视频分段建模？

\author{Zhiyu He}
\email{hezy22@mails.tsinghua.edu.cn}
\orcid{0000-0003-1291-2739}
\affiliation{%
  \institution{DCST, Tsinghua University}
  \institution{Quan Cheng Laboratory}
  % \institution{Zhongguancun Laboratory}
  \city{Beijing}
  \country{China}
}

\author{Zhixin Ling}
\email{lingzhixin@kuaishou.com}
\orcid{0000-0003-0967-2287}
\affiliation{%
  \institution{Kuaishou Technology}
  \city{Beijing}
  \country{China}
}

\author{Jiayu Li}
\orcid{0000-0002-6351-897X}
\email{jy-li20@mails.tsinghua.edu.cn}
% \email{lijiayu997@gmail.com}
\affiliation{%
  \institution{DCST, Tsinghua University}
  \city{Beijing}
  \country{China}
}

\author{Zhiqiang Guo}
\orcid{0000-0001-9393-4854}
\email{georgeguo.gzq.cn@gmail.com}
\affiliation{%
  \institution{DCST, Tsinghua University}
  \city{Beijing}
  \country{China}
}

\author{Weizhi Ma}
\authornote{Corresponding Authors.\label{correspond}}
\orcid{0000-0001-5604-7527}
\email{mawz@tsinghua.edu.cn}
\affiliation{%
  \institution{AIR, Tsinghua University}
  \city{Beijing}
  \country{China}
}

\author{Xinchen Luo}
\email{luoxinchen@kuaishou.com}
\orcid{0009-0008-7261-5297}
\affiliation{%
  \institution{Kuaishou Technology}
  \city{Beijing}
  \country{China}
}

\author{Min Zhang\textsuperscript{\ref{correspond}}}
\orcid{0000-0003-3158-1920}
\email{z-m@tsinghua.edu.cn}
\affiliation{%
  \institution{DCST, Tsinghua University}
  \institution{Quan Cheng Laboratory}
  \city{Beijing}
  \country{China}
}

\author{Guorui Zhou}
\orcid{0009-0002-8550-279X}
\email{zhouguorui@kuaishou.com}
\affiliation{%
  \institution{Kuaishou Technology}
  \city{Beijing}
  \country{China}
}

\renewcommand{\shortauthors}{Zhiyu He et al.}

%% file: Header/2.abstract.tex
%%
%% The abstract is a short summary of the work to be presented in the
%% article.
\begin{abstract}

% Short videos have become highly popular, necessitating the development of effective recommender systems that match users with content they are interested in. 
The rapid growth of short videos has necessitated effective recommender systems to match users with content tailored to their evolving preferences. 
Current video recommendation models primarily treat each video as a whole, overlooking the dynamic nature of user preferences with specific video segments. In contrast, our research focuses on segment-level user interest modeling, which is crucial for understanding how users' preferences evolve during video browsing. 
To capture users' dynamic segment interests, we propose an innovative model that integrates a hybrid representation module, a multi-modal user-video encoder, and a segment interest decoder. 
% to capture users’ dynamic interests. 
Our model addresses the challenges of capturing dynamic interest patterns, missing segment-level labels, and fusing different modalities, achieving precise segment-level interest prediction.

We present two downstream tasks to evaluate the effectiveness of our segment interest modeling approach: video-skip prediction and short video recommendation. Our experiments on real-world short video datasets with diverse modalities show promising results on both tasks. 
It demonstrates that segment-level interest modeling brings a deep understanding of user engagement and enhances video recommendations. 
We also release a unique dataset that includes segment-level video data and diverse user behaviors, enabling further research in segment-level interest modeling. 
This work pioneers a novel perspective on understanding user segment-level preference, offering the potential for more personalized and engaging short video experiences. % By focusing on video segment's user interest, our method improves recommendation accuracy and has practical applications, such as personalized content discovery and optimized video creation strategies.

\end{abstract}

%% file: Header/3.ccs_keywords.tex
\ccsdesc[500]{Information systems~Recommender systems}
\ccsdesc[500]{Information systems~Personalization}

\keywords{
Video recommendation, User modeling, Segment-level interest.}

%%
%% Keywords. The author(s) should pick words that accurately describe
%% the work being presented. Separate the keywords with commas.
% \keywords{Music recommendation. Ubiquitous computing...}

%% file: Main/1.Introduction.tex
\section{Introduction}
\label{sec:intro}

% Outline: (core: segment-level interest is important!!!) 
% 区别：（1）whole->segment, (2) explicit
% 挑战：（1）无直接标签 （2）行为-兴趣差异、id-模态差异

\begin{figure}[t]
    \centering
    \setlength{\abovecaptionskip}{0.0cm}
    \setlength{\belowcaptionskip}{-0.3cm}
    \includegraphics[width=8.8cm]{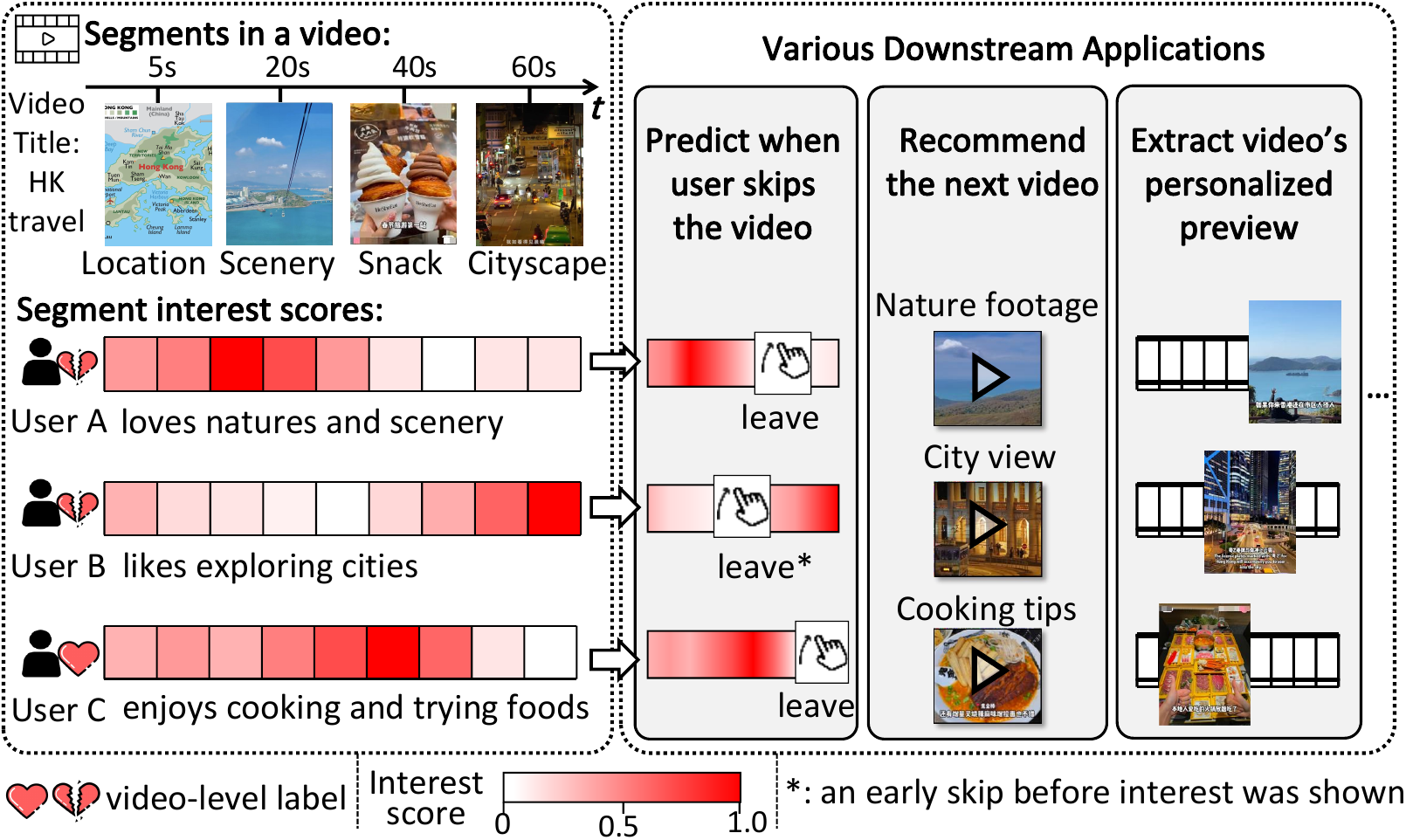}
    \caption{Users' dynamic interests in video segments reflect their diverse preferences, offering deeper insight than the overall video preference. % It supports 
    Such interests benefit 
    downstream applications such as video-skip prediction, recommendations, and personalized homepage thumbnails. 
    % Take one video as an example. 
    }
    \label{fig:motivation}
\end{figure}

% 1-1 video recommendation:短视频越来越流行; the need for recommendation
Short videos have gained immense popularity in recent years, raising the need for effective recommender systems to match users with content they may interested in. 
% 1-2 the phenomenon -- segment interest:（+过渡句）浏览视频时，用户喜欢不同的片段，会表现为滑走等时间轴行为 
Unlike traditional media such as movies and magazines, short videos are characterized by rapid pacing and frequent scene transitions~\cite{li2024narrative,opara2025impact}. 
This results in highly dynamic user interests during video consumption, manifested by shifting attention and preferences across different segments. 
These preference shifts are often reflected in behaviors along the video timeline, such as skipping to the next video, which should be captured by models for better recommendations. 
% Unlike traditional media consumption, users' engagements with short videos are highly dynamic, often influenced by the contents of specific segments in the videos. %Their interest in a video changes as they browse, a pattern that is often reflected in timeline behaviors, such as skipping or pausing.
% Users' interest in a video changes dynamically during browsing. 
% As they favor different video segments, this is often reflected in behaviors along the timeline, such as skipping to the next video.

% 2 introduce multi-modality information(是否base多模态？）: regard video as a whole entity. this way loss the features on time sequence 
% 现有短视频推荐的研究通常把视频作为整体，这种方式损失了视频内时间序列的特征，难以与用户时间轴的行为（例如播放时间）匹配。
However, segment-level preferences have not been explored well in previous studies, as most existing works treat videos as whole entities, extracting features and training id embeddings to assign an overall preference score~\cite{chen2018temporal,gu2024holistic}.
When incorporating multimodal information, researchers typically focus on content features from the video cover or fixed frames~\cite{wei2019mmgcn,jing2024multimodal,chen2024multi}, overlooking the dynamic temporal nature within videos.
Preference label is typically user feedback at video-level, such as ``skip'' or ``effective view'',% without considering segment-level behavioral signals to discover dynamic preferences.
while segment-level behavioral signals are severely ignored. 
The absence of segment-level information fundamentally limits the recommender system's accuracy. % better user understanding 
\citet{shang2023learning} steps further by exploring detailed visual data along the video's timeline. They constructed a graph representing user preferences implicitly by selecting positive or negative frames separately based on predefined rules.
However, in real-world scenarios, video segments are tightly coupled instead of being isolated entities, as the preceding and following segments influence the user's interest in a specific segment. 
The separated selection overlooks the interest's continuity, resulting in an underdeveloped user interest modeling. 

% 4 引到显式预测频段兴趣很重要
% 4-1 While browsing video, interest changes over time.
%浏览视频时，随着视频内用户喜欢不同的片段，用户兴趣会在一个视频中积累，通过建模用户在视频内时间维度的兴趣分布，能增强视频表征，指导短视频推荐
As shown in the left part of Figure~\ref{fig:motivation}, a short video typically comprises segments with different topics and rapid transitions, and users' interests evolve when they engage in short video browsing. 
Different segments appeal to different users based on their personalized preferences. 
% Some segments might capture a user’s attention more effectively, while others might be less engaging. 
% While traditional recommendation models fail to identify interest dynamics; we argue that segment-level interest modeling is crucial for improving recommendation accuracy and user satisfaction. 
Modeling this personalized segment-level interest aware of the temporal transition, which was neglected in previous models, is crucial for short video recommender systems.
% It provides a deeper understanding of user preferences, improving recommendation accuracy and user satisfaction. 
% Explicitly modeling users' segment-level interests along the temporal dimension allows us to capture the dynamic interest distribution within a video. 
% This provides a deeper understanding of user preferences, thereby improving the accuracy of short video recommendation systems.
Segment-level interest modeling offers significant value in practice, as illustrated in the right part of Figure~\ref{fig:motivation}. 
% video-skip prediction
First is \textbf{video-skip prediction}, which identifies low-interest segments users will likely skip. 
% video recommendation
It also improves \textbf{video recommendation} by incorporating segment-level interest scores, which capture deeper preferences than traditional video-level approaches. 
Additionally, \textbf{personalized thumbnails} in the homepage can be generated by selecting high-interest segments, driving users to click on the video and watch.
% 4-2 Protential values:
% 同时具有的潜在价值：（1）推荐个性化的缩略视频片段（首页）；（2）模拟视频受欢迎片段，指导创作者
% For example, it enables the recommendation of personalized thumbnail video segments on homepages, making content discovery more engaging. Besides, it can simulate the popularity of specific video segments, offering valuable insights to creators for optimizing content creation.
% Potential applications also include artificially personalized video editing.
% simulating the popularity of specific video segments to guide content creation and enabling personalized video editing, where low-interest parts are fast-forwarded or removed for a more tailored viewing experience. 
These encourage a comprehensive study on segment-level interest modeling for essential user understanding, targeting considerable improvement in recommendation accuracy and user satisfaction. 

% 5-1 Goal
Thus, we aim to model the users' interests in short video segments for personalized recommendations. 
However, modeling user segment interests is challenging:
% 5-1 challenges: (1)interest 随时间变化 (2)没有兴趣的直接标签（3） (3)modality gap - 对齐
(1) User interest evolves along the video timeline, influenced by the human a patterns of human attention, which are independent of video content. Effectively incorporating these patterns to capture interest dynamics is a key challenge. 
% (1) Human inherent temporal attention pattern. This attention pattern, sometimes decoupled to video segment content, keeps influencing user dynamic interest along the video timeline. Effectively incorporating such influence to capture interest dynamics presents a key challenge. 
(2) Segment-level user feedback is typically implicit and sparse, such as scrolling actions~\cite{lin2024inverse}, providing little information on user interest for each video segment. This lack of explicit labels complicates the task of modeling segment interest. 
% (2) The lack of segment-level user labels. Segment-level user feedback is typically implicit sparse signals such as scrolling actions~\cite{lin2024inverse}, providing little information on user interest for each video segment and thus complicating the model's training and evaluation procedure.  
 (3) Modeling segment interests needs video content information, and the multi-modal fusion challenge arises when combining the complementary signals from user-item interactions and content-based information. 
% The gap between collaborative filtering signals from user-item interactions and content-based information extracted from video segments makes their seamless fusion challenging. 
% Our method
To alleviate the challenges, we propose an innovative model to capture segment-level user interest, comprising a hybrid representation, a multi-modal user-video encoder, and a segment interest decoder. It serves inner-video segment position indices as embedding input and fits position bias for challenge~1. 
%Hybrid representation represents videos and users from collaborative filtering signals and multi-modal content features. Multi-modal user-video encoder leverages representation from various modalities to model the intricate interactions between users and video content, providing modal-specific segment interests. Segment interest decoder includes modality fusion and intra-video position bias, to learn the segment interest score in timeline. 
For challenge~2, we design an intra-video loss function to exploit implicit relationships between user feedback and segment interests, addressing feedback sparsity and guiding the training process. 
Our model represents and encodes interactions within each modality and performs late-stage fusion to ensure modality-specific interests while allowing insights from all modalities to complement each other~(challenge~3). 
%bridge the modality gap

% 6-1 tasks
We present two downstream tasks to evaluate the effectiveness of our interest modeling in the recommendation. The first task is \textit{video-skip prediction}, where we predict when or which segments a user is likely to skip based on their fine-grained interests.
The second task is \textit{video recommendation}, which integrates segment-level interest predictions to enhance the overall performance of video recommendations. 
% 6-2 experiements and results
The experimental results on both tasks demonstrate the accuracy of segment interest modeling and its value in recommendation. 
We present cases to highlight the effectiveness and potential value of interest modeling, providing a valuable perspective on segment interest modeling in recommendations. 
We also release a \textit{video recommendation dataset} that includes segment-level video data and diverse user behaviors, which is the first dataset to provide both of these critical components.

The contributions of this work are as follows:

\vspace{-0.2em}
% \begin{itemize}
\begin{itemize}[nolistsep, leftmargin=*]
    \item  To the best of our knowledge, we are the first to address the problem of segment-level user interest modeling in short video recommendation, extending the existing paradigm from video-level to segment-level modeling. This helps better user understanding and provides a more personalized experience.
    %To our knowledge, we are the first to model segment-level user interests in short video recommendations. 
    % We capture the rapid dynamics of video content and user interests,  extending the existing paradigm from video-level to segment-level modeling.
    
    \item We propose a novel segment-level user interest model with hybrid representation, multi-modal user-video encoder, and segment interest decoder, tackling challenges of capturing dynamic interest, modality gaps, and missing direct segment-level user feedback. % by integrating inter-video position bias, modality fusion, and designing an intra-video loss function to exploit implicit relationships between user feedback and segment interests.
    
    \item  Experiments on a public dataset and a new dataset show encouraging results of the new paradigm and model on two tasks: video-skip prediction and video recommendation. We also release the dataset with segment-level video data and diverse user behavior data, which is the first data source for segment-level user preference understanding~\footnote{Codes and data are available at https://github.com/hezy18/SegMMInterest/\label{foot:link}}.
    % offering a new perspective and data source for segment-level user preference understanding. 
    % in video recommendation, paving the way for more engaging and personalized experiences on short video platforms. 

\end{itemize}

%% file: Main/2.Related_work.tex
\section{Related Work}
\label{sec:related}

\subsection{Fine-grained User Modeling} 

% 区分fine-grained 的概念，同样是从整体到细粒度，但如何细粒度是不同的：
% 几种不同的fine-grained
% (1) multi-dimension interest
% (2) temporal unit-level interest

% 用户兴趣建模是推荐系统中的一个常见问题，研究者利用成分的信息学习用户兴趣表示。我们的工作是从视频整体聚焦到细粒度视频片段，进行细粒度的用户建模。我们将现有细粒度用户建模的工作概括为两类，modeling multi-dimension interest 和 temporal unit-level interest。
%User interest modeling is a common topic in recommendation systems, where researchers leverage various components of information to learn user interest representations. 
% Our work shifts the focus from holistic video modeling to fine-grained segment modeling, exploring fine-grained user modeling in greater detail. Existing works in this area can be categorized into two main types: multi-dimension interest and temporal unit-level interest. 
Our work shifts the focus from holistic video modeling to fine-grained segment modeling, with a deeper exploration of fine-grained user modeling. These related works can be categorized into two types: multi-dimensional interest and temporal unit-level interest.

% multi-dimension interest 将对item的兴趣细分为对多个类别、多层级的兴趣。一部分工作将用户历史交互的item聚类成多个类别，根据user item的交互或用户之间的共现，刻画用户对不同类别的兴趣\cite{tian2022multi，du2024disentangled，jiang2020aspect}。另一部分工作考虑不同行为反馈信号（clik，buy，forward）等，学习用户不同方面或不同反馈的偏好\cite{meng2023coarse}。这些方法一般是将粗粒度建模模块和细粒度建模模块结合，增强在商品层级的推荐效果。
Multi-dimensional interest breaks down user interest in an item into multiple categories or hierarchical levels. Some studies cluster items based on user interaction history and model interests across different categories~\cite{tian2022multi, du2024disentangled, jiang2020aspect}. Others examine varying behavioral feedback signals, such as clicks, purchases, and shares, to capture user preferences across different types of interactions~\cite{meng2023coarse}. These methods often combine coarse-grained and fine-grained modeling to improve recommendation accuracy.

% temporal unit-level interest 将用户兴趣表示根据时序建模。一些工作考虑短期兴趣（以此区分与长期历史兴趣表示）\cite{meng2024coarse}或进行在历史上多时间尺度的兴趣建模，另一部分将item根据其内容的时间关系分成更小的单元，建模对单元的兴趣。这类工作多数为news recommendation。用户会点击新闻中的词语，包含对该词感兴趣的细粒度信息。因此可以建模 token level interest，结合ducoment level interest 推荐新闻 \cite{qi2022fum, hou2023multi}
Temporal unit-level interest focuses on modeling user interests based on fine-grained unit level. These researches mainly target news recommendations, where users' interest is represented by their engagement with individual tokens within an article, such as specific words or phrases. By combining token-level interest with document-level interest, these models aim to improve the relevance of news recommendations~\cite{qi2022fum, hou2023multi}.
% 与第二类一致，我们将segment作为unit，但新闻推荐利用用户点击特定word的反馈，我们是短视频推荐场景，利用用户离开片段的反馈。
Similar to the second approach, we treat segments as the basic unit in our model. However, while news recommendation systems utilize user clicks on specific words as feedback, our work focuses on short video recommendations, where user feedback is based on the exit from segments, bringing a unique challenge to our work.

\subsection{Multi-Modal Video Recommendation}  
%介绍MMRec
% difference： 整体->分段

% Multi-modal video recommendation systems leverage a large amount of multimedia content information, alongside user interaction data, to enhance personalization recommendation~\cite{liu2024multimodal}. The handling of multi-modal data in video recommendation can be broadly categorized into fusion-based and graph-based methods.
% Fusion-based methods combine these signals into unified representations using techniques like attention or bilinear pooling. For instance, UVCAN~\cite{liu2019user} uses self-attention to fuse user-side and item-side multimodal data. M3CSR~\cite{chen2024multi} clusters IDs and uses a modal encoder to integrate features from different modalities for cold-start video recommendation. 
% Graph-based approaches, like MMGCN~\cite{wei2019mmgcn} and MMGCL~\cite{yi2022multi}, model user-item interactions through bipartite graphs, while MMKGV~\cite{liu2022multi} uses a graph attention network to integrate multimodal information via knowledge graphs. 
% These methods aim to capture complex relationships between users and items for more accurate recommendations. 

Our proposed model utilizes both multimedia content and user interaction data for fine-grained segment-level user interest modeling, making it a multi-modal system.
Generally, multi-modal recommender systems 
% Multi-modal video recommender systems use both multimedia content and user interaction data to improve personalization~\cite{liu2024multimodal}. These systems 
adopt fusion-based or graph-based methods for handling multi-modal data~\cite{liu2024multimodal}. Fusion-based methods combine signals into unified representations using techniques like attention or bilinear pooling. For example, UVCAN~\cite{liu2019user} employs self-attention to merge user-side and item-side data, while M3CSR~\cite{chen2024multi} integrates features from multiple modalities for cold-start recommendations. Graph-based methods, such as MMGCN~\cite{wei2019mmgcn} and MMGCL~\cite{yi2022multi}, model user-item interactions with bipartite graphs, and MMKGV~\cite{liu2022multi} uses a graph attention network for knowledge graph-based fusion. These approaches aim to capture complex relationships between users and items.

However, most existing works treat videos as a whole, focusing on video-level features for recommendations. Only a few address the inherent diversity within a video. For instance, \citet{shang2023learning} models positive and negative frames for video-level recommendations but overlooks temporal continuity between segments. In contrast, our approach explicitly models segment-level user interest,  taking into account the temporal relationships between segments. 
Besides, \citet{chen2017personalized} focus on key-frame recommendations based on time-synchronized user feedback. However, our work examines user interest across different segments of a video involving downstream applications. 
We propose a novel model to explore how users’ interests vary across different segments in short video recommendations, as well as the broader downstream applications of video-skip prediction and video-level recommendations.

\subsection{Video Highlight Detection} 

% CV中用到视频细粒度的类似topic是“Personalized Highlight Detection” （ICCV， MM...)
% 从图像处理的角度，根据用户历史视频预测未来视频的高光检测点
% 和我们工作根本上差异：任务不同，且并未用到CF
%（CF也不是推荐的必须，所以可以考虑做baseline）

Video highlight detection~(VHD) is a topic in Computer Vision related to our work, as both divide video into segments and aim to capture content patterns in temporal sequences. 
The primary objective of VHD is retrieving a subset of video frames that capture a person’s primary attention from the original video~\cite{rochan2020adaptive,wei2022learning}. 
PHD-GIFs~\cite{garcia2018phd} is the first personalized VHD technique that extracts highlight predictions guided by subjects' manually created GIFs. % 强调用户注释
Models are constructed by convolutional layers, fully connected layers, and learned parameters based on individuals' preferred clips to guide a content-based highlight detection network~\cite{rochan2020adaptive,chen2021pr}.  
% \cite{panagiotakis2020personalized}
\citet{bhattacharya2022show} builds upon this approach by employing a multi-head attention mechanism, achieving improved performance. 

However, modeling segment-level user interests has fundamental differences from VHD. 
VHD only uses image features without user behavior as model input, while interest modeling is based on collaborative information, in which user interaction is combined with videos for personalized and comprehensive understanding. 
Besides, user-uninvolved human-selected labels (GIFs) form the datasets for VHD training and evaluation, while in real-world short video recommendation scenarios, it's difficult to explicitly figure out users' "chosen segments", which is discussed in Section~\ref{sec:intro} as challenge~2. 
As a result, further discussion or comparison between VHD and our proposed model is biased and not objective, thus is not included in the remainder of this paper. 
% 然而，这类方法完全依赖于图像特征之间的相似性，标签为用户手动挑选的GIFs，用户之间并无重复的商品交互。然而在真实短视频的推荐场景中，并没有用户preferred clip这一标签，而只有用户离开位置。另外，我们最终应用到下游的推荐场景，充分利用用户商品的协同关系十分重要。
% However, these methods rely entirely on the similarity between image features, with labels based on user-selected GIFs, and there is no overlap in product interactions between users. In contrast, in real-world short video recommendation scenarios, there is no explicit label for a user’s "preferred clip". Furthermore, applying our model to downstream recommendation tasks requires collaborative relationships between users and items.  
% 没有源代码（show me what i like这篇等几篇，基于图像网络，对比价值不大，因此没有作为baseline对比
% 可否用needs

% 区别1：任务目标的差异。P-VHD only needs to predict the label of frames to highlight or no-highlight one ； 而我们的工作对每个片段都进行建模 (改成说标签的差异）
% 区别2：方法的差异。用CNN的方法，寻找被试标注过的高亮片段和目标检测视频中相似的图像，并没有用到用户交互和协同的信息。而我们基于推荐的场景，用户商品的协同十分重要。

%% file: Main/3.Method.tex
% \section{Notations and Problem Definition}
% \label{sec:notation}

\section{User Segment Interest Modeling}
\label{sec:method}
% model命名：
% FINE: Fine-grained Interest Network for video (Encoding)不好
% SPOT: Segment-based Personalized Orientational (Transformer)不好
% SegIN：In同时具有interest和In video的含义 
% SegInterest

\begin{figure*}[htbp]
    \centering
    \setlength{\abovecaptionskip}{0.cm}
    \includegraphics[width=18cm]{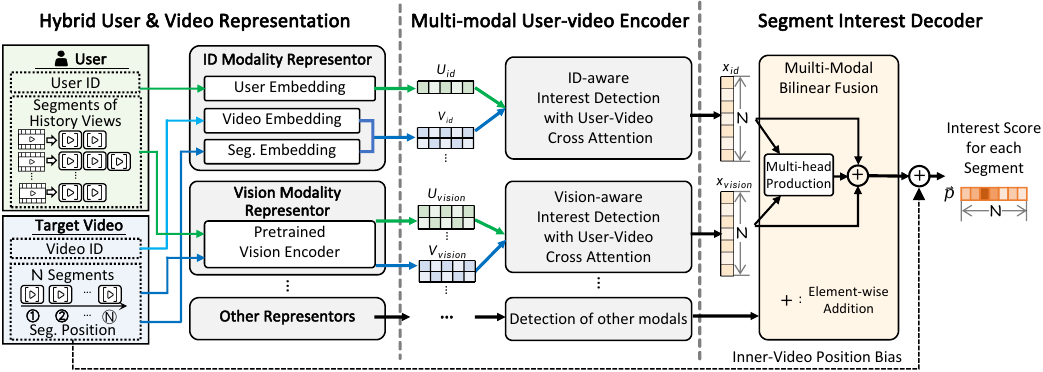}
    \caption{Overview of user segment interest modeling with hybrid user and video representation, multi-modal user-video encoder, and segment interest decoder. $N$ is the number of segments in the target video, and segment interest scores are the model's output.}
    \label{fig:model} 
\end{figure*}

\subsection{Problem Definition}

On short video platforms, users continuously watch videos lasting a few seconds to a few minutes. Videos are shown one at a time, and users swipe up or down to view the next system-recommended video.
Formally, $\mathcal{U} = \{u_1, u_2, ...\} $ denotes the set of users, and $\mathcal{V} = \{v_1,v_2,...\}$ represents the set of items~(i.e., videos). 

\textbf{Video segments} are defined by evenly dividing videos with the same time intervals: Each video $v \in \mathcal{V}$ is divided into $N$ segments, $S_{v} = (S_{v}^1, S_{v}^2, ..., S_{v}^N)$, where the duration of $S_v^i$ is $t$ for any $i$.
% We divide each video $v \in \mathcal{V}$ into $N$ segments, $S_{v} = (S_{v}^1, S_{v}^2, ..., S_{v}^N)$, with evenly spaced time intervals. 
% Interactions with fast-forwarding operations are excluded, 
The viewed segments are determined with the viewing time of each video,
and we assume that users watch videos from the beginning, as this behavior is common in video streaming.
% We determine the viewed segments based on the viewing time of each video.
We define the \textbf{segment interest modeling} problem in the context of sequential recommendation, where the user's history of viewed videos is denoted as $(v_1,v_2,...)$.  
The corresponding user history of $u$ viewed segments is denoted as $S_{u} = (S_{u}^1,S_{u}^2,...,S_{u}^M)$, where $M$ is the number of viewed segments. 
Given a user $v$ and a target video $v$, our goal is to infer the interest score sequences $\vec{p} = (p_{1},p_{1},...p_{N})$ for video segments $S_{v}$, where $p_{i}$ is the interest score for segment $S_{v}^i$. 
These segment interest scores can be used for downstream applications.

% \subsection{Segment Interest Model}

\subsection{Model Overview}

We propose a model for user segment interest with Hybrid User \& video Representation, Multi-modal User-video Encoder, Segment Interest Decoder, as shown in Figure~\ref{fig:model}. 

Hybrid User-video Representation learns embeddings from different modalities for the user and target video. The ID modality is represented through embedding layers, while feature encoders process other modal inputs, such as visual features. 
The representations from the user and video interact independently for each modality in the Multi-modal User-video Encoder. We design modal-aware interest detection with user-video cross-attention. It outputs modal-specific interest scores from the interaction information of each modality. 
Finally, the Segment Interest Decoder fuses different modalities' outputs and generates segment interest scores. 

 % 值得强调的是，模型具有模态上的泛化能力，用同样的方式，该模型可增加或减少模态。
Notably, the model is modality-agnostic; it can easily accommodate changes in the number or type of modalities. 

\subsection{Hybrid User \& video Representation}
\label{sec:representer}

User and video representations are derived from  inputs across different modalities. This  work represents users and videos using two modalities: the visual modality and the ID modality. 
The visual modality captures content-related features, and this representation process can be extended to other modalities, such as auditory and textual. The ID modality, on the other hand, consists of user and video identifiers, enabling personalization.

As for the visual modality, we leverage CLIP~\footnote{https://huggingface.co/openai/clip-vit-large-patch14-336}~\cite{radford2021learning} as the pre-trained Vision Encoder to extract feature representations. These features are then mapped to the desired representation space through a linear projector, resulting in $U_{vision}$ and $V_{vision}$, derived from the user's viewed segments $S_{u}$ and target video segments $S_{v}$ inputs, respectively. 
For ID modality, we first embed the user ID to obtain the user embedding $U_{id}$. 
To distinguish between different segments within the same video, we introduce segment positions ranging from 1 to $N$. We then embed the video ID and the segment positions. The video ID's embedding and segment position embeddings are concatenated to form the video segment representations, $V_{id}$. 

This design strikes a balance between content-driven semantics and unique identifiers, allowing the model to effectively capture both contextual and personalized information.

\subsection{Multi-modal User-video Encoder} 
\label{sec:encoder}

% representations输入->得到 N相关的
The Multi-modal User-video Encoder includes Modal-aware Interest Detection to achieve segment-related interest representations by processing user-video interactions within each modality separately. 

\begin{figure}[t]
    \centering
    \setlength{\abovecaptionskip}{0.cm}
    \setlength{\belowcaptionskip}{-0.3cm}
    \includegraphics[width=8.8cm]{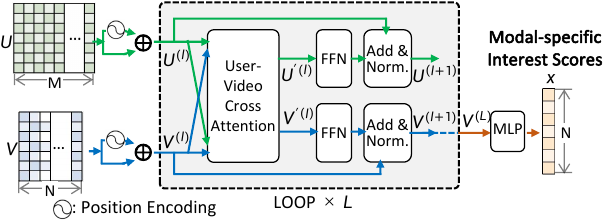}
    \caption{Details of the interest detection module in the multi-modal encoder. $U$ and $V$ denote the representations of the user and target video.} %M>N in vision modality, M=1 in ID modality. 
    %N represents the number of segments in target videos. 
    \label{fig:model_detection} 
\end{figure}

% 解释为什么要specifically design: 
Figure~\ref{fig:model_detection} demonstrates Modal-aware Interest Detection. 
The core module of Modal-aware Interest Detection is User-Video Cross-attention. 
Unlike conventional transformer-based methods that treat users and items uniformly within a shared sequence, we recognize that user and item representations possess distinct syntactic and semantic structures.   
Inspired by~\citet{wang2023ms}, we design the User-Video Cross-Attention. 
For simplicity, we denote $U_{vision}$ and $U_{id}$ as $U$, and $V_{vision}$ and $V_{id}$ as $V$. 
In the vision modality,$\quad V \in \mathbb{R}^{N \times d}$ and $U \in \mathbb{R}^{M \times d}$, while in ID modality,$\quad U\in \mathbb{R}^{1 \times d}$. 
Due to the use of the user's historical sequence, the number of segments viewed by the user is greater than the number of segments in the target video ($M$ > $N$).  

We first encode sequence positions to incorporate the order of elements. 
% Then $U^1$ and $V^1$ enter the loop, iterating for $L$ layers. 
Then, the model enters a loop with
each layer contains a User-video Cross-attention. % and a Feedforward Network~(FFN) to transform the representations, with residual connections and layer normalization applied throughout. 
The User-video Cross-attention aggregates the user embeddings $U^{(l)}$ and video embeddings $V^{(l)}$ to form user-interated video representations, $V'^{(l)}$, through the following process:

\begin{equation}
{A}_{VU}^{(l)} = \mathit{FFN}(V^{(l)}) \cdot \mathit{FFN}(U^{(l)})/\sqrt{d} \label{eq:A_VU}
\end{equation}

\begin{equation}
{A}_{VV}^{(l)} = \mathit{FFN}(V^{(l)}) \cdot \mathit{FFN}(V^{(l)})/\sqrt{d} \label{eq:A_VV}
\end{equation}

\begin{equation}
{V'}^{(l)} = \mathrm{Softmax}({A}_{VU}^{(l)} \oplus {A}_{VV}^{(l)}) \cdot \left(\mathit{FFN}(V^{(l)}) \oplus FFN(U^{(l)})\right) \label{eq:V^l+1}
\end{equation} % \oplus \bigoplus \|
where $\oplus$ represents the concatenation operation. A similar set of equations is used to obtain ${U'}^{(l)}$: 

\begin{equation}
{A}_{UU}^{(l)} = \mathit{FFN}(U^{(l)}) \cdot \mathit{FFN}(U^{(l)})/\sqrt{d} %\label{eq:A_UU}
\end{equation}

\begin{equation}
{U'}^{(l)} = \mathrm{Softmax}({A}_{VU}^{(l)} \oplus {A}_{UU}^{(l)}) \cdot \left(\mathit{FFN}(V^{(l)}) \oplus \mathit{FFN}(U^{(l)})\right) \label{eq:V^U+1}
\end{equation}

Through the Feedforward Network~(FFN), residual connections, and layer normalization, we obtain $U^{(l+1)}$ and $V^{(l+1)}$ from ${U'}^{(l)}$ and ${V'}^{(l)}$.
$U^{l+1}$ and $V^{l+1}$ are inputs to the next layer. We adopt $L$ layers to aggregate the user embedding. The encoder output, $V^{(L)}$, represents video segment-related latent scores, aggregated from user representations. 
We then apply an MLP~(Multilayer Perceptron) module to reduce the representations from high to low dimensions, resulting in segment interest scores $x$ from different modalities. 
% Here, Conv1D is a non-overlapping 1D convolution with stride and kernel size set to R.

\subsection{Segment Interest Decoder}
\label{sec:decoder}

% After obtaining the latent scores $\{x_1, x_2, ...\}$ of the user-interacted video segments from different modalities, we focus on decoding them into the final segment interest scores. 

% As $\{x_1, x_2, ...\}$ come from different modalities, use multi-modal bilinear fusion to preserve their individual strengths and enable them to complement each other. 

The segment interest decoder receives the latent scores of the user-interacted video segments to decode them into the final interest score.
Since those scores result from various modalities, multi-modal bilinear fusion is adapted to preserve their separate strengths while enabling complement integration among each other as: 

% \begin{equation}
% o = \sigma\left(b_{f} + \mathit{Proj}({x}_{1}) + \mathit{Proj}({x}_{2}) + \mathit{Proj}({x}_{1}^\top{x}_{2})\right)  \label{eq:fusion}
% \end{equation} 
\begin{equation}
o = \sigma\left(b_{f} + \sum_i \mathit{Proj}({x}_{i}) + \sum _{i \not = j} \mathit{Proj}({x}_{i}^\top{x}_{j})\right)  \label{eq:fusion}
\end{equation} 

where $x_i$ represents the output from different Modal-aware Interest Detection, $\mathit{Proj}(\cdot)$ denotes independently-trained Linear Projectors and $b_{f}$ is a trainable bias term. 
${x}_{i}^\top{x}_{j}$ refers to the multi-head product of the modality-specific features, which enables capturing diverse interactions while enhancing computational efficiency. 
% If the linear weights of the ${x}_{i}^\top{x}_{j}$ become zero metrics, the fusion process degenerates to a traditional concatenation followed by a linear layer. 
% The model can flexibly handle single or multiple modalities. It applies to additional modalities by adding individual terms and pairwise combinations. 

To model the human inherent temporal attention pattern, for example, 
user interest typically accumulates or decreases as they watch a video from start to finish, and the probability of interest decreases for later segments, we introduce an inner-video position bias to capture the relationship between segment position and interest.  
This position bias is added to the fusion output $o$, producing the final segment interest score $\vec{p}$:

\begin{equation}
\vec{p} = o + (w_{p} \cdot \vec{pos} + b_{p})  \label{eq:bias}
\end{equation} 
where $w_{p}$ and $b_{p}$ are learnable parameters for position bias, $\vec{pos} = (idx_{1},idx_{2},...idx_{N})$. $\vec{p} = (p_{1},p_{1},...p_{N})$ represents the interest scores for the segments, with $p_i$ being the interest score for segment $S_v^i$.

\subsection{Training and Inference} 
\label{sec:training}
% Especially the inter-video BPR loss

% Training

% 在数据和真实场景中，我们没有用户对于片段层级的兴趣标签。因此，我们利用用户滑走所处片段这一反馈设计训练的损失函数。
As discussed in Section~\ref{sec:intro}, it's unaffordable to have user feedback for every video segment. 
In real-world applications, the only practical user interaction is the skipped segment position, during which he/she gets bored or distracted and skips to the next video. 
% 由于片段兴趣体现在同一视频中不同片段之间的差别，我们计算intra-video loss.
% Since segment interests are reflected in the differences between segments within the same video, we compute an intra-video loss.
Since segment interests are reflected in the differences between segments within the same video, a carefully designed intra-video loss function based on the skipped segment position is proposed to fully exploit the limited user interaction within video watching.
% 对于用户对每一个片段$S_v^i$的兴趣分数$p_i$, 分数越高，用户继续浏览的可能性越高。尽管可能用户因为过早的离开导致一些位置之后即使有喜欢的也看不到，但用户离开的位置一定是相对兴趣更低的。
% For each segment $S_v^i$, the interest score $p_i$ indicates the likelihood that the user continues watching. 
% A user always leave after watching part of a video, resulting in no feedback for the remaining segments. We assume that the segment where they left should have a lower interest score compared to the segments they have already watched. The loss function is defined as:
Specifically, the loss function assumes that the skipped segment position should have the lowest interest level among the watched segments, with the loss for one video segment pair $\left(u, v\right)$:

\begin{equation}
\mathcal{L}_{u,v} = -\sum\nolimits_{(i = y,j\neq y)}^{N-1} ln \sigma(p_j - p_i)  \label{eq:loss_uv}
\end{equation} 

where interest score $p_i$ indicates the likelihood of continuing watching for the user, $y$ denotes the skipping segments of user $u$ toward target video $v$, and $N$ is the segment number. 

The overall loss is defined as the average of $\mathcal{L}_{u,v}$ across user-video interactions in the training set: 

\begin{equation}
\mathcal{L} = \frac{1}{|\mathcal{D}|} \sum_{(u,v)\in \mathcal{D}} \mathcal{L}_{u,v}  \label{eq:loss_all}
\end{equation} 

During inference, the model directly computes the interest scores for each segment in the target video.

\subsection{Potential Application for Downstream Tasks}

% or: To summarize,

% discussion: 作为 (Sec3和Sec4之间的衔接）
% 得到的segment interest score包含了对用户及视频内容理解的丰富信息，与推荐紧密联系。因此可以有丰富的下游应用，包括增强视频粒度的推荐表现，包括自动剪辑视频的片段作为部分提供个性化兴趣的预告，包括在视频发布前提示创作者视频不同片段对用户的吸引程度。但遗憾的是，这其中的很多任务受限于离线数据的缺失，没有办法评估。在受限的数据中，我们提出两个任务，用于评估用户兴趣分数的准确性和有效性。
The segment interest scores obtained from our model encapsulate rich information about both user preferences and video content, closely tied to recommendation systems. These scores open up various downstream applications, such as enhancing video-level recommendation performance, video automatical editing and thumbnail generation. %. enabling automatic video segment generation to showcase personalized video thumbnails, and providing insights to creators on the appeal of different video segments before release. %However, many of these tasks are constrained by the lack of offline data, making evaluation difficult. 
% Here, we propose two tasks to assess the accuracy and effectiveness of the user interest scores: video-skip prediction~(Section~\ref{sec:task1}) and short video recommendation~(Section~\ref{sec:task1}).  
To illustrate the effectiveness of our proposed user interest scores, we conduct two tasks:
(1) Video-skip prediction~(Section~\ref{sec:task1}) to directly validate the accuracy of user interest scores, and (2) Short video recommendation~(Section~\ref{sec:task2}) as a common and important task to evaluate whether segment interest scores are useful in real-world scenarios.
% Two tasks are proposed to assess the accuracy and effectiveness of the user interest scores: video-skip prediction~(Section~\ref{sec:task1}) and short video recommendation~(Section~\ref{sec:task2})
% Specifically, video-skip prediction verifies the segment users skip or leave the video, attempting to deeply characterize user interest and video information in the recommender system, whose results can serve as a supplementary metric in online recommendations.
%预测用户在哪里跳过，用于在推荐系统中对用户和视频内容进行更深的刻画，反映用户兴趣深度，从而可以用作联合优化指标辅助推荐决策。
% (2) Short video Recommendation~(Section~\ref{sec:task2}): deciding whether to push video to specified users, the fundamental task in the recommendation realm. 

%% file: Main/4.Task1.tex
\section{Video-skip Prediction~(Task~1)
}
\label{sec:task1}

\subsection{Task Description and Settings}

\subsubsection{Task Description}

% This task predicts when a user will likely skip a video. As a downstream application of segment-level user dynamic interest modeling, we predict the video skipping position based on segment-level interest scores~(outputs of Section~\ref{sec:method}). We rank the segments within a video according to their interest scores. Higher interest indicates a lower likelihood of skipping. 
The video-skip prediction task predicts which segment users will possibly skip the video. It attempts to characterize user interest and video content in the recommender system, serving as a supplementary metric for online recommendations. 
% is a downstream application of our proposed segment-level interest modeling,
After the generation of segment-level interest scores~(Section~\ref{sec:method} outputs), segments are ranked, as the scores are negatively correlated to the likelihood of skipping. 
% For each user-video interaction, HR@K and NDCG@K are assessed by judging if the actual skipped segment appears in the top-K segments (we set $K = 1, 5, 10$) and computing the corresponding Normalized Discounted Cumulative Gain, respectively. 
For each user-video interaction, we assess whether the actual skipped segment appears in the top-K segments~(HR@K) and compute the corresponding Normalized Discounted Cumulative Gain~(N@K). We set K=1,5,10.

% Video-skip prediction verifies the segment users skip or leave the video, attempting to deeply characterize user interest and video information in the recommender system, whose results can serve as a supplementary metric in online recommendations.

\subsubsection{Dataset}

% 解释数据集的选用，也强调SegMM数据的优势
% 目前有哪些短视频数据集，其他的为什么不行
% 根据我们的任务和目标，短视频推荐数据集中需要包含用户对视频的时间戳滑走行为反馈；同时，最好还能有片段粒度的模态特征（视觉、听觉、文本）来体现视频内容。目前在公开数据集中，并没有包含这两种信息的数据。MicroLens~\cite{}是很好的数据集，包含raw video，但是并没有用户对视频的滑走及视频时间戳行为。因此，我们公开一个数据集SegMM包含细粒度的特征和丰富的用户时间戳反馈信号。同时，我们以KuaiRand作为代表，展示只用D模态的表现。
Based on our task objectives, the short video recommendation dataset we used must include user timestamp feedback on videos~(watch time or skipping). It is preferable to have segment-level modality features (visual, auditory, and textual) to capture video content. 
Currently, no public dataset contains both types of information. Here, we release a dataset~(SegMM) to fill this gap. 
Some public datasets include ID and behavioral features. Considering the differences in time span and the number of interactions compared to SegMM, we select the KuaiRand dataset. 

\textbf{SegMM}\textsuperscript{\ref{foot:link}} is a dataset from a commercial short video platform. It contains 2,369 randomly sampled active users with millions of interactions over 3 days~(June 01 to June 03, 2024). It has visual features for each segment of raw videos and the behaviors of these users' interaction. To our knowledge, it's the first short video recommendation dataset with both fine-grained features and behaviors. 
% \textsuperscript{\ref{foot:link}}.

\textbf{KuaiRand}~\cite{gao2022kuairand} is collected from one popular Chinese short video app, KuaiShou. We use the data of recommended items. It contains 983 users and their interactions with system-recommended short videos over two weeks~(April 07 to April 21, 2022).

% Note that we considered MicroLens~\cite{ni2023content} as it's a valuable dataset with raw videos; however, it lacks user timeline behavior data. % such as watch time. 

\begin{table}[htbp]
\small
\renewcommand\arraystretch{1.1}
\setlength{\abovecaptionskip}{0.cm}
\setlength{\belowcaptionskip}{0.cm}
\caption{Datasets statistics. ``Visual'' refers to whether the data provides visual features.}  % view 这些也可以解释一下%``\#Segment'' and ``Visual''  refer to the number of segments and whether the data provides visual features, respectively.
\centering
\begin{tabular}{cccccc @{}}
\hline
\textbf{Datasets} & \textbf{\#Users} &
\textbf{\#Items} & \textbf{\#View} & \textbf{\#Segment} & \textbf{Visual} \\
\hline
% EEGSVRec & 30 &  2,636 & 3,657 &  & \checkmark \\
SegMM & 2,369 & 362,430 & 902,115 & 3,920,483 & \checkmark \\
KuaiRand & 983 & 717,652 & 1,615,315 & 8,140,477 & -- \\

\hline
\end{tabular}
\label{tab:dataset}
\vspace{-1.5em}
\end{table}

% Interactions with fast-forwarding operations are excluded to % 保证我们的假设：用户by default从头观看，watch time 转化成skip position。
% Interactions with fast-forwarding operations are either excluded in dataset annotation or too few to be safely omitted for SegMM and KuaiRand, respectively. 
Interactions with fast-forwarding operations are excluded in SegMM.
% For both datasets, interactions with fast-forwarding operations are excluded. 
Therefore, we can assume that videos are displayed from the beginning, and the skip position of each video is directly derived from users' watch time.
% Therefore, assuming that videos are displayed from the beginning, user watch time is directly treated as skipping position. 
We partitioned the data %into training, validation, and testing sets 
using an 8:1:1 user-based split. Short videos typically range from a few seconds to several minutes. To facilitate task evaluation, we selected videos no longer than 200 seconds and assumed each segment to be 5 seconds. 
The statistics of datasets are shown in Table~\ref{tab:dataset}.

\subsubsection{Baselines}

\begin{table*}[htbp] 
\small
\caption{The video-skip prediction performance. $^{*}$ and $^{**}$ indicate p-value<0.1 and <0.05 from t-test between \textit{Ours} and the best baseline. \textbf{Bold} and \underline{underline} show the best of all results and the best baseline results. N is short for NDCG.}
\centering

\setlength{\abovecaptionskip}{-0.5cm}
\setlength{\belowcaptionskip}{-1cm}
\renewcommand{\arraystretch}{1.1}
\begin{tabular}{@{}p{1.4cm}p{1.3cm}|p{1cm}<{\centering}p{1cm}<{\centering}p{1cm}<{\centering}p{1cm}<{\centering}p{1cm}<{\centering}|p{1cm}<{\centering}p{1cm}<{\centering}p{1cm}<{\centering}p{1cm}<{\centering}p{1cm}<{\centering}@{}} % 
\hline
 & Dataset & \multicolumn{5}{c|}{\textbf{SegMM}} & \multicolumn{5}{c}{\textbf{KuaiRand}}\\
 \cline{2-12}
Category & Methods & HR@1 & HR@5 & N@5 & HR@10 & N@10 & HR@1 & HR@5 & N@5 & HR@10 & N@10\\
\hline
\multirow{4}{*}{MostPopular} & Random & 0.0244 & 0.1258 & 0.0737 & 0.2530 & 0.1142 & 0.0251 & 0.1266 & 0.0746 & 0.2515 & 0.1145 \\
& AllPosition & 0.1309 & 0.2625 & 0.2007 & 0.3679 & 0.2343 & 0.0897 & 0.2111 & 0.1523 & 0.3242 & 0.1884\\
& UserPosition & 0.1504 & 0.2823 & 0.2204 & 0.3850 & 0.2531 & 0.0992 & 0.2219 & 0.1629 & 0.3336 & 0.1985 \\
& ItemPosition & \underline{0.2619} & 0.3824 & 0.3269 & 0.4701 & 0.3548 & \underline{0.2648} & 0.3790 & 0.3299 & 0.4669 & 0.3580 \\
\hline
\multirow{4}{*}{GeneralRec} & LightGCN & 0.1148 & 0.2380 & 0.1808 & 0.2734 & 0.1924 & 0.0866 & 0.2199 & 0.1567 & 0.2803 & 0.1762 \\
& DirectAU & 0.1684 & 0.3680 & 0.2770 & 0.4128 & 0.2917 & 0.1263 & 0.3284 & 0.2334 & 0.4065 & 0.2587 \\
& Caser & 0.1655 & 0.3742 & 0.2782 & 0.4296 & 0.2963 & 0.1216 & 0.3091 & 0.2207 & 0.3816 & 0.2443 \\
& SASRec & 0.1831 & 0.4381 & 0.3201 & 0.5048 & 0.3420 & 0.1554 & 0.3940 & 0.2821 & 0.4839 & 0.3112 \\
\hline
\multirow{7}{*}{ContextRec} & WideDeep & 0.1306 & 0.2765 & 0.2095 & 0.3118 & 0.2211 & 0.1231 & 0.3404 & 0.2381 & 0.4311 & 0.2675 \\
& DCNv2 & 0.1525 & 0.3518 & 0.2605 & 0.4054 & 0.2780 & 0.1420 & 0.4419 & 0.2997 & 0.5665 & 0.3401\\
& FinalMLP & 0.1663 & 0.4032 & 0.2941 & 0.4683 & 0.3154 & 0.1465 & 0.4336 & 0.2979 & 0.5571 & 0.3379\\
& AdaGIN & 0.2177 & \underline{0.5868} & \underline{0.4167} & \underline{0.6778} & \underline{0.4465} & 0.1597 & 0.4376 & 0.3066 & 0.5539 & 0.3442\\
& DIN & 0.1312 & 0.4030 & 0.2744 & 0.4988 & 0.3055 & 0.1542 & 0.4177 & 0.2934 & 0.5259 & 0.3286 \\
& DIEN & 0.1622 & 0.4118 & 0.2955 & 0.4920 & 0.3217 & 0.1812 & \underline{0.4535} & \underline{0.3247} & \underline{0.5702} & \underline{0.3625} \\
& CAN & 0.2031 & 0.4817 & 0.3517 & 0.5640 & 0.3785 & 0.1762 & 0.4280 & 0.3095 & 0.5302 & 0.3426 \\
\hline
% \multirow{4}{*}{MMRec} & MMGCN & 0.1311 & 0.4325 & 0.2899 & 0.5628 & 0.3321 & \multicolumn{5}{c}{\multirow{4}{*}{------}}\\
\multirow{3}{*}{MMRec} & SLMRec & 0.1136 & 0.2333 & 0.1773 & 0.2745 & 0.1908 & \multicolumn{5}{c}{\multirow{3}{*}{(no visual data)}} \\
& BM3 & 0.1506 & 0.3639 & 0.2641 & 0.4387 & 0.2886 & \\
& FREEDOM & 0.1808 & 0.3783 & 0.2874 & 0.4243 & 0.3025 & \\
\hline
% \multirow{3}{*}{\parbox{1.4cm}{WatchTime\\Prediction}} & WLR & 0.1436 & \multicolumn{4}{c|}{\multirow{3}{*}{(N/A)}} & 0.0468 & \multicolumn{4}{c}{\multirow{3}{*}{(N/A)}} \\ % can not produce that results
% & D2Q & 0.1227 & & & & & 0.1321 & & & & \\
% & TPM & \underline{0.3049} & & & & & 0.1663 & & & & \\
% \hline
% $^{**}$
\multirow{3}{*}{Ours} & ID & 0.3353$^{**}$ & 0.7676$^{**}$ & 0.5462$^{**}$ & 0.8534$^{**}$ & 0.5740$^{**}$ & 0.2904  & 0.5709 & 0.4275 & 0.7378 & 0.4818 \\
%\textbf{0.3689} & \textbf{0.5734} & \textbf{0.4718} & \textbf{0.8142} & \textbf{0.5485}
& Visual & 0.3828$^{**}$ & 0.8171$^{**}$ & 0.6186$^{**}$ & 0.8567$^{**}$ & 0.6318$^{**}$ & \multicolumn{5}{c}{(no visual data)}\\
& Both & \textbf{0.4072}$^{**}$ & \textbf{0.8214}$^{**}$ & \textbf{0.6228}$^{**}$ & \textbf{0.9225}$^{**}$ & \textbf{0.6572}$^{**}$ & \textbf{0.2904}$^{*}$  & \textbf{0.5709}$^{**}$ & \textbf{0.4275}$^{**}$ & \textbf{0.7378}$^{**}$ & \textbf{0.4818}$^{**}$\\
\hline
\end{tabular}
\label{tab:task1_results}
\end{table*}

Four types of baselines are employed in this task. 
Intuitively, the most popular method of calculating position probabilities is applied. 
Since few prior works concerned the video skip prediction task, three traditional recommenders~(GeneralRec, ContextRec, MMRec) are implemented by generating prediction scores for each segment and ranking those scores inside videos for final evaluation. 
% Furthermore, watch-time prediction is also compared due to its similarity to the target task. 
Follows detail those baselines: 
% For our baselines, we employ the most popular methods that account for position bias in video skipping, recommender algorithms with segment-level features, and watch-time prediction models due to their task similarity. 
% 关于为什么要用这些baseline
% （1）the most popular
% （2）目前没有直接做video skip prediction的方法，我们用recommender algorithms对每一segment预测得到分数，然后在video内对这些分数ranking，从而evaluate
% （3）Watch-time prediction和我们的任务（video-skip prediction）可以相近，因此，我们页对比了这类方法

\begin{enumerate}[nolistsep, leftmargin=*]
\item \textbf{MostPopular} uses skip probabilities for each segment position, derived from statistics on the training and validation datasets. It includes random selection, overall position probability~(AllPosition), user-specific position probability~(UserPosition), and item-specific position probability~(ItemPosition). 
\item \textbf{GeneralRec} represents standard recommendation methods employing collaborative filtering techniques such as LightGCN~\cite{he2020lightgcn} and DirectAU~\cite{wang2022towards}. Additionally, it includes sequence-based recommenders like Caser~\cite{tang2018personalized} and SASRec~\cite{kang2018self}.
\item \textbf{ContextRec} incorporates contextual information into recommendations. We includes classic models like WideDeep~\cite{cheng2016wide} and DCNv2~\cite{wang2021dcn}, as well as state-of-the-art methods such as FinalMLP~\cite{mao2023finalmlp} and AdaGIN~\cite{sang2024adagin}. It also considers historical context with models like DIN~\cite{zhou2018deep}, DIEN~\cite{zhou2019deep}, and CAN~\cite{bian2022can}. 
\item \textbf{MMRec} encompasses multi-modal recommendation methods that integrate various data modalities, including SLMRec~\cite{tao2022self}, BM3~\cite{zhou2023bootstrap}, and FREEDOM~\cite{zhou2023tale}. % MMGCN~\cite{wei2019mmgcn}
% \item \textbf{Watch-time Prediction Methods} including ~WLR~\cite{covington2016deep} by positive sample weighting, D2Q~\cite{zhan2022deconfounding} by group regression, and TPM~\cite{lin2023tree} as a representative work of and tree-based classification methods. % 解释只能在HR@1上评价
\end{enumerate}

% WLR：Deep Neural Networks for YouTube Recommendations（RecSys16）：正负样本预估，播放时长做加权的交叉熵loss，inference为视频时长*正样本概率
% D2Q：Deconfounding Duration Bias in watch time Prediction for Video Recommendation（KDD22）：MSE优化观看百分比的分位点，根据视频时长所在的组别的统计值，得到该分位点对应的播放时长
% TPM：Tree based Progressive Regression Model for watch time Prediction in Short video Recommendation（KDD23）：对播放时长的分位点分桶，将预测分位点变成树的结构，得到观看时长落在每个叶子节点区间的概率，交叉熵loss优化；inference：概率加权的各区间的平均时长之和

\subsubsection{Implementation Details}

We implement recommender methods using ReChorus~\cite{wang2020make,li2024rechorus2} and multi-modal recommender using MMRec~\cite{zhou2023mmrec}. 
% The baselines of the Watch time prediction follow the original paper's code. Similar to the TPM method, proportion normalization of the interest scores' reciprocal is used as probability weight as the result of watch time when computing MAE.
We implement our method with PyTorch\footnote{https://pytorch.org/}. 
The model is optimized by Adam~\cite{kingma2014adam} optimizer with tunable learning rate and embedding size, where the batch size is 1024. Learning rate are tuned from 1e-4 to 1e-2 and embedding size are from 32 to 128. % Layers are tuned in {[64], [128], [64,64]}. 
As for our model and the baselines that consider history, we set max history as 20.
% In our experiments, we use the setting of K = 1, 3, 5, 10. 
We use NDCG@5 on the valid set for early stopping if the performance does not increase in 10 epochs.  
% Our code and detailed hyperparameter settings are available\textsuperscript{\ref{foot:link}}.

\subsection{Overall Video-skip Ranking Performance}
\label{sec:Task1results}

Table~\ref{tab:task1_results}
% and Table~\ref{tab:task1_WatchTime} 
show the performances of our proposed method compared to baselines. 
%（1）相比于baseline，我们的方法在两个数据集上显著提升，验证片段兴趣分数建模的准确性, 说明具有video-skip prediction任务的能力
Our method demonstrates significant improvements over baselines across two datasets, validating the accuracy of segment interest modeling. We highlight its effectiveness in capturing user interests at the segment level to boost video-skip prediction.

% \begin{table}[t] 
% \small
% \caption{The performance comparison between our model and baselines on the watch-time prediction.}
% \centering
% \setlength{\abovecaptionskip}{-0.3cm}
% \setlength{\belowcaptionskip}{-0.3cm}
% \renewcommand{\arraystretch}{1.2}
% \begin{tabular}{@{}p{0.9cm}p{0.7cm}|p{0.7cm}<{\centering}p{0.7cm}<{\centering}p{0.7cm}<{\centering}|p{0.7cm}<{\centering}p{0.7cm}<{\centering}p{0.7cm}<{\centering}@{}} % 
% \hline
% Dataset & Metrics & WLR & D2Q & TPM & \multicolumn{3}{c}{Ours} \\
%  & & & & & ID & Visual & Both\\
% \hline
% \multirow{2}{*}{SegMM} & HR@1 & 0.1436 & 0.1227 & \underline{0.3049} & 0.3353$^{**}$ & 0.3828$^{**}$ & \textbf{0.4072}$^{**}$ \\
%  & MAE~$\downarrow$ & 3.3790 & \underline{3.1228} & 3.3217 & 3.0254$^{*}$ & 3.0399$^{*}$ & \textbf{2.9564}$^{**}$\\
%  % \hline
% \multirow{2}{*}{KuaiRand} & HR@1 & 0.0564 & 0.1321 & \underline{0.1663} & \textbf{0.2904}$^{**}$ &  \multicolumn{2}{c}{\multirow{2}{*}{(no visual data)}}\\
%  & MAE~$\downarrow$ & 4.5244 & \underline{3.9337} & 4.2714 & \textbf{3.5940}$^{**}$ & \\
% \hline
% \multicolumn{8}{l}{\parbox{8cm}{MAE~(Mean Absolute Error) is a typical measure of regression accuracy in watch time prediction tasks.}}
% \end{tabular}
% \label{tab:task1_WatchTime}
% \end{table}

%（2）可以看出片段兴趣受到位置的影响，item- specific的positon probability的效果较好，尤其是在HR@1上，说明用户skip在同一视频上的一致性。
% 分析不同类别的baseline
Analyzing the performance of different baseline categories, we find that
% The analysis reveals that 
segment interests are notably influenced by position within the video. Among the most popular strategies, item-specific position probability yields superior performance, particularly in the HR@1 metric. This suggests a high consistency in user skipping behavior within the same video. %underscoring the importance of incorporating item-specific positional information in recommendation models.
% 还要解释为什么部分方法超不过item pos 甚至超不过 user pos 乃至 all pos. 部分recommendation method 表现比most popular还要差，例如LightGCN。尽管输入包含了segment所处的位置，这些方法并不是为了检测片段兴趣而提出的，模型并没有特别处理这部分信息。
Some recommendation methods, such as LightGCN, perform even worse than the most popular category since they are not specified for segment interest and cannot characterize the segment information.
% （3）在对segment recommend 分数并排序兴趣时，SegMM最好的baseline是AdaGIN，KuaiRand最好的baseline是CAN，context-aware methods在这个任务中具有优势。
% （4）MMRec类别方法中，即使引入position，但这个特征的贡献难以被模型挖掘，仅靠视觉模态特征效果仍旧不足。
When evaluating segment recommendation scores and interest rankings, AdaGIN emerges as the top-performing baseline for the SegMM dataset, while CAN leads for the KuaiRand dataset, indicating that context-aware methods hold a distinct advantage in this task. For the MMRec category, BM3 and FREEDOM perform better than their backbone (LightGCN) by including visual features. %even when positional features are introduced, their contribution remains limited, and reliance solely on visual modality features does not suffice.

% For Table~\ref{tab:task1_WatchTime}, the performance comparison highlights the ability of our model to estimate watch time. 
% （5）watch time prediction拟合观看时长，通过回归的指标评价差异，而在准确性上，较我们的方法仍有所差距。
% 在watch time prediction task上，我们的方法比该任务的代表性？经典？方法要好。即使都在只用ID的情况. % due to 我们将用户历史行为和目标视频的在片段粒度上的交叉
% MAE指标计算预测watch time和真实watch time之间的距离，而HR@1衡量watch time是否处于为真实watch time。我们的方法在计算MAE时用兴趣分数的倒数比例归一化（proportion normalization）作为权重/概率，得到播放时长（watch time）的预测，baseline例如TPM也是类似的策略。
% In watch time prediction tasks, our proposed method outperforms typical algorithms %under fair comparison by only applying ID information
% , which may attributed to the segment-level cross-attention between user history behavior and the target video.  % 
% % MAE represents the distance between prediction and ground truth, while HR@1 measures the accuracy. Similar to the TPM method, proportion normalization of the interest scores' reciprocal is used as probability weight when computing MAE, resulting in the prediction of watch time.

% 分析运行时间的成倍差别
% \textcolor{red}{analyze the time cost differences between ours and Recommender. The time cost of Recommender is the \#segment per video times of Our time.}兴趣建模独立于下游任务，带来实际应用的潜力。
% Considering efficiency, traditional recommenders output interest scores for each segment separately, while our method computes segment-level serial interests with the whole video as input, significantly reducing the time cost of the system by \#segment times, building the potential for more flexible real-world scenarios. % Interest modeling is decoupled with downstream applications

% 总结句：
In summary, our method achieves remarkable performance in video-skip prediction, a critical subtask within recommendation systems. By accurately modeling segment-level user interests, our approach effectively identifies points of lowest user engagement where skips are most likely to occur. The experimental results underscore the superiority of our method, ensuring more reliable and personalized video recommendations.

\subsection{Video-skip Ranking Performance on Cold-start Items}
\label{sec:Task1cold}

% item pos的影响
% 多模态
% 冷启视频的效果

\begin{table}[htbp] 
\small
\caption{Video-skip prediction performance on cold videos.}
\centering
\setlength{\abovecaptionskip}{-0.3cm}
\setlength{\belowcaptionskip}{-0.5cm}
\renewcommand{\arraystretch}{1.1}
\begin{tabular}{@{}p{1.3cm}|p{0.8cm}<{\centering}p{0.8cm}<{\centering}p{0.8cm}<{\centering}|p{0.8cm}<{\centering}p{0.8cm}<{\centering}p{0.8cm}<{\centering}@{}} % 
\hline
 Dataset & \multicolumn{3}{c|}{\textbf{SegMM}} & \multicolumn{3}{c}{\textbf{KuaiRand}}\\
 \hline
Methods & H@1 & H@5 & H@10 & H@1 & H@5 & H@10 \\
\hline
AllPosition & 0.1395 & 0.2764 & 0.3795 & 0.0887 & 0.2109 & 0.3229 \\
UserPosition & 0.1672 & 0.3001 & 0.3959 & 0.1040 & 0.2291 & 0.3382 \\
\hline
DirectAU & 0.1649 & 0.4015 & 0.4639 & 0.1302 & 0.3587 & 0.4454\\
SASRec & 0.1916 & 0.4553 & 0.5334 & 0.1790 & 0.4169 & 0.5023\\
AdaGIN & \underline{0.2045} & \underline{0.5621} & \underline{0.6642} & 0.1623 & 0.4496 & 0.5693 \\
DIEN & 0.1719 & 0.4586 & 0.5493 & \underline{0.1940} & \underline{0.4693} & \underline{0.5700}\\
% CAN & 0.2091 & 0.4986 & 0.5815 & 0.1748 & 0.4439 & 0.5538 \\
FREEDOM & 0.1812 & 0.3801 & 0.4220 & \multicolumn{3}{c}{(no visual data)} \\
% TPM & & \multicolumn{2}{c|}{(N/A)} & & \multicolumn{2}{c}{(N/A)} \\
\hline
Ours(ID) & 0.3211 & 0.7638 & 0.8496 & 0.2662 & 0.5582 & 0.7308 \\
%0.3735 & 0.5953 & 0.8254 \\
Ours(Visual) & 0.4060 & 0.8073 & 0.8519 & \multicolumn{3}{c}{(no visual data)}\\
Ours(Both) & \textbf{0.4068} & \textbf{0.8170} & \textbf{0.9234} & \textbf{0.2662} & \textbf{0.5582} & \textbf{0.7308} \\
\hline
\end{tabular}
\label{tab:task1_cold}
\end{table}

% item specific position probability在HR@1相较其他baseline有较好的表现，且比用全局的position probability好很多，展示出同一个视频在不同用户滑走的位置上有一定的一致性。我们的方法也引入片段所处的position，因此也学习到了不同视频的更可能滑走的位置。为验证方法在用新视频上的表现，我们将模型在训练和验证集中未出现的视频进行验证。cold item 数量约占1/3。Table~\ref{tab:task1_cold} shows the results on cold items. Appendix 中展示了非cold item上的表现（Table~\ref{tab:task1_noncold}）。

Utilizing item-specific position probability~(ItemPosition) significantly outperforming the global position probability~(AllPosition), demonstrating consistency in the skipping positions in the same video. 
% Our method also incorporates the segment position and multi-modal contents, %, allowing it to learn the more likely interaction positions for different videos. 
It raises concerns about the performance of cold videos. %, we tested it on items that did not appear in the training and validation sets. 
Approximately one-third of the general test set is with cold video, and their performance is shown in Table~\ref{tab:task1_cold}.

% 实验发现在cold 和非cold上的表现，我们的方法仍相比baseline有显著提升。对于cold item，推荐的模型均超过most popular的方法，体现出推荐方法对video理解的泛化能力。最好的baseline是？？和？？，xxx
The experimental results indicate that the recommender methods consistently outperform the most popular approach for cold items, and our method still significantly outperforms the baselines on both cold and non-cold items, reflecting the model's generalization capability in understanding videos. 
% 注意到SegMM上冷item表现更好，KuaiRans上冷item表现更差，这是由于...（TODO！）
% \textcolor{red}{Whether to explain why cold is better than all (especially on SegMM. -> generally, cold-item has been skipped earlier (by statistics), amplify bias. }
% \textcolor{red}{Explain the cold vs original: id dows not worse on cold because we still learns the position and the user pattern, which may not harm the performance a lot
% . Image and Both are the same as the original's results because the features from multi-modality brings video content understanding.}比较Table2 和Table4 仅有ID输入时cold-item效果略差，有content信息后cold item效果较好， 在cold item上，各方法的表现并不会差很多，这是由于使用了片段所处位置的信息，对于cold item，模型并非一无所知，
% As shown in Table~\ref{tab:task1_cold} and Table~\ref{tab:task1_results}, the methods didn't perform well for cold items with only ID input but achieved better measurements by adding content information. The performance of those algorithms didn't show noticeable differences in cold items due to the usage of segment position information, preventing the modal from being innocent to cold items. 
Incorporating segment position information into our model ensures that it is not entirely unaware of cold videos. 
Comparing Table~\ref{tab:task1_cold} and Table~\ref{tab:task1_results}, our proposed model performs slightly worse on cold videos when only using ID input. The performance on cold videos improves by incorporating content information~(Visual modality), demonstrating its generalization capability in understanding videos.

\subsection{Ablation Study for Video-skip Prediction}
\label{sec:Task1ablation}

 \begin{figure}[htbp]
    \centering
    \setlength{\abovecaptionskip}{0.0cm}
    \setlength{\belowcaptionskip}{-0.3cm}
    \includegraphics[width=8.6cm]{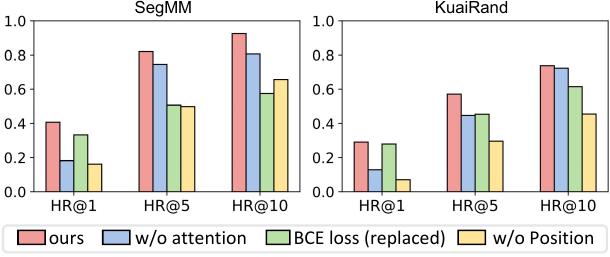}
    \caption{Ablation study of attention mechanism, segment position indices, and replacing our loss with BCE~(Binary Cross-Entropy) loss.}
    \label{fig:ablation} 
\end{figure}

We conduct ablation studies on the input modality, interest detection module, segment position, and intra-video loss. 

% input modality
% （1）我们的方法可以将多个模态相结合，因此我们在只用collaborative information~(ID modality）和仅visual modality都进行了实验。如Table~\ref{tab:task1_results}所示。我们发现visual modality比ID modality表现更好。visual modality利用了用户历史浏览片段和目标视频片段的内容，二者之间建立的encoder可以很好的捕捉它们之间的关联，进一步使visual modality有更好的表现，由于bilinear fusion的作用，ID和vision的结合表现最好。说明了ID和vision modaliy利用了不同含义的特征，可以相互补充
Our method fuses multiple modalities, combining collaborative (ID) and visual information. As shown in Table~\ref{tab:task1_results}, the visual modality outperforms the ID modality. This is because the visual modality leverages both user history and target video content, with the encoder effectively capturing their relationships. Bilinear fusion of modalities yields the best performance, highlighting the complementary nature of different features.

% segment position
%（2）在视频拆分时，每一个segment对应在视频中的position,并应用于segment embedding 和 inner-video position bias. 其相对位置也通过position encoding引入cross attention。因此，我们实验去掉position信息之后的结果。Figure~\ref{fig:ablation} 展示了结果，与完整方法下降很多，展示了模型能充分利用segment position。
We also remove cross-attention in the interest detection module and segment position indices. The attention module captures user-video relationships, while segment position indices are integrated into segment embeddings and intra-video position bias.
% Each segment has a specific position in the video, incorporated into segment embedding and inner-video position bias. 
Figure~\ref{fig:ablation} shows performance dropped when they were removed, highlighting the importance of user-video cross-attention and segment position for the model. Especially for KuaiRand, where there's no visual data, segment positions are crucial for distinguishing between segments. 
% interest detection module
% we designed user-video cross-attention in interest detection module to capture the connection between user representations and video representations. To validate its ability, we variant user-video cross-attention to self-attention and MLP with cross user and video representations. 如Figure~\ref{fig:ablation}所示，Self-attention 损失很多性能，因为只利用到video representations。 变体实验体现了user-video cross-attention的重要性。

Finally, we replaced our intra-video loss with BCE~(binary cross-entropy) loss. BCE treats each segment independently and assumes skipped positions have an interest label of 0, causing a significant performance drop. This highlights the effectiveness of our method in addressing challenges with missing segment-level labels. 

%% file: Main/5.Task2.tex
% \vspace{-1em}
\section{Short Video Recommendation(Task~2)} 
\label{sec:task2}

% \subsection{Task Description}

\subsection{Segment-integrated Video Recommendation Framework}

 \begin{figure}[htbp]
    \centering
    \setlength{\abovecaptionskip}{0.cm}
    \setlength{\belowcaptionskip}{0.cm}
    \includegraphics[width=8.5cm]{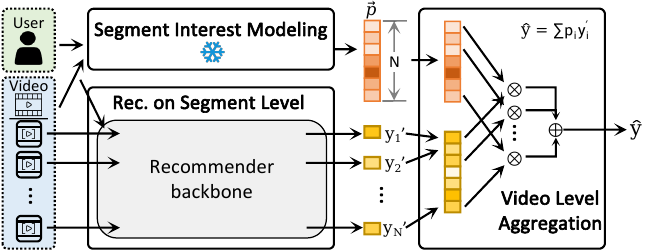}
    \caption{SegRec: Segment-integrated Video Recommendation Framework. The parameter of segment interest modeling is frozen, serving segment interest scores $\vec{p}=(p_1,p_2,..p_N)$ to down-streaming video recommendation.}
    \label{fig:task2_framework} 
\end{figure}

\begin{table*}[htbp] 
\small
\caption{The video-level recommendation performance. \textbf{Bold} and \underline{underline} show the best of all results and the best baseline results. $^{*}$/$^{**}$ indicates p-value<0.1/0.05 from the t-test conducted over five repetitions. }
\centering
\setlength{\abovecaptionskip}{0.0cm}
\setlength{\belowcaptionskip}{-0.5cm}
\renewcommand{\arraystretch}{1.1}
\begin{tabular}{@{}p{1.1cm}p{1.3cm}|p{0.9cm}<{\centering}p{0.9cm}<{\centering}p{0.9cm}<{\centering}|p{0.9cm}<{\centering}p{0.9cm}<{\centering}p{0.9cm}<{\centering}|p{0.9cm}<{\centering}p{0.9cm}<{\centering}p{0.9cm}<{\centering}|p{0.9cm}<{\centering}p{0.9cm}<{\centering}p{0.9cm}<{\centering}@{}} % 
\hline
\multicolumn{14}{c}{\textbf{SegMM}}\\
\hline
 % & Dataset & \multicolumn{6}{c|}{\textbf{SegMM}} & \multicolumn{6}{c}{\textbf{KuaiRand}}\\
 % \hline 
 & Backbone & \multicolumn{3}{c|}{\textbf{WideDeep}} & \multicolumn{3}{c|}{\textbf{AdaGIN}} & \multicolumn{3}{c|}{\textbf{DIN}} & \multicolumn{3}{c}{\textbf{CAN}} \\
Group & Methods & AUC & F1 & Logloss$\downarrow$ & AUC & F1 & Logloss$\downarrow$ & AUC & F1 & Logloss$\downarrow$ & AUC & F1 & Logloss$\downarrow$\\
\hline
\multirow{3}{*}{Self-Info} & Video & 0.7301 & 0.6550 & 0.6188 & 0.7371 & 0.6634 & 0.6032 & 0.7471 & 0.6809 & 0.5992 & 0.7502 & 0.6901 & 0.5927 \\
& SegSum & 0.7310 & 0.6597 & 0.6185 & 0.7347 & 0.6669 & 0.6078 & 0.7438 & 0.6748 & 0.6068 & 0.7495 & 0.6861 & 0.5931 \\
& SegAdjust & 0.7312 & \underline{0.6682} & 0.6154 & 0.7382 & 0.6709 & 0.5978 & \underline{0.7489} & 0.6808 & 0.6006 & \underline{0.7539} & \underline{0.6926} & \underline{0.5887} \\
\hline
\multirow{5}{*}{\parbox{1.1cm}{Weight from Task1 baseline$^\dag$}} & ItemPosition & 0.7329 & 0.6588 & 0.6163 & 0.7418 & 0.6671 & 0.6082 & 0.7433 & 0.6680 & 0.5998 & 0.7423 & 0.6392 & 0.6048 \\
 & Backbone & 0.7295 & 0.6517 & 0.6151 & \underline{0.7421} & \underline{0.6833} & \underline{0.5968} &  0.7426 & 0.6602 & 0.6009 & 0.7388 & 0.6901 & 0.6092 \\
 & SASRec & 0.7288 & 0.6503 & 0.6244 & 0.7298 & 0.6689 & 0.6182 & 0.7358 & 0.6623 & 0.6074 & 0.7419 & 0.6595 & 0.6022 \\
& AdaGIN & \underline{0.7338} & 0.6612 & \underline{0.6085} & \underline{0.7421} & \underline{0.6833} & \underline{0.5968} & 0.7464 & \underline{0.6812} & \underline{0.5971} & 0.7494 & 0.6890 & 0.5932 \\
% \multicolumn{3}{c|}{Same as ``Backbone''}
& FREEDOM & 0.7322 & 0.6568 & 0.6108 & 0.7412 & 0.6720 & 0.6003 & 0.7409 & 0.6799 & 0.5995 & 0.7445 & 0.6852 & 0.5975 \\
\hline
\multirow{3}{*}{\textbf{SegRec}} & ID & 0.7439$^{**}$ & 0.6742$^{**}$ & 0.5963$^{**}$ & 0.7443$^{*}$ & 0.6883$^{*}$ & 0.5957 & 0.\textbf{7584}$^{**}$ & 0.6910$^{**}$ & 0.5834$^{**}$ & \textbf{0.7579}$^{*}$ & 0.6949 & \textbf{0.5846}$^{*}$ \\
& Vision & 0.7419$^{**}$ & 0.6725$^{*}$ & 0.5970$^{*}$ & 0.7424 & 0.6853 & 0.5972 & 0.7580$^{**}$ & 0.6904$^{**}$ & 0.5837$^{**}$ & 0.7557 & 0.6921 & 0.5865 \\
& Both & \textbf{0.7441}$^{**}$ & \textbf{0.6762}$^{**}$ & \textbf{0.5959}$^{**}$ & \textbf{0.7452}$^{*}$ & \textbf{0.6919}$^{*}$ & \textbf{0.5945} & 0.7581$^{**}$ & \textbf{0.6915}$^{**}$ & \textbf{0.5827}$^{**}$ & 0.7558 & \textbf{0.6997} & 0.5875 \\
\hline
\hline
\multicolumn{14}{c}{\textbf{KuaiRand}}\\
\hline
 & Backbone & \multicolumn{3}{c|}{\textbf{WideDeep}} & \multicolumn{3}{c|}{\textbf{AdaGIN}} & \multicolumn{3}{c|}{\textbf{DIN}} & \multicolumn{3}{c}{\textbf{CAN}} \\
Category & Methods & AUC & F1 & Logloss$\downarrow$ & AUC & F1 & Logloss$\downarrow$ & AUC & F1 & Logloss$\downarrow$ & AUC & F1 & Logloss$\downarrow$\\
\hline
\multirow{3}{*}{Self-Info} & Video & \underline{0.7214} & \underline{0.6449} & \underline{0.6133} & 0.7313 & 0.6616 & 0.6059 & 0.7347 & 0.6697 & 0.6039 & 0.7450 & 0.6752 & 0.5953\\
& SegSum & 0.7027 & 0.6383 & 0.6363 & 0.7301 & 0.6606 & 0.6071 & 0.7283 & 0.6677 & 0.6086 & 0.7447 & 0.6787 & 0.5951 \\
& SegAdjust & 0.7073 & 0.6403 & 0.6342 & \underline{0.7318} & \underline{0.6629} & \underline{0.6031} & 0.7325 & \underline{0.6760} & 0.6059 & \underline{0.7454} & \underline{0.6792} & \underline{0.5950} \\
\hline
\multirow{4}{*}{\parbox{1.1cm}{Weight from Task1 baseline$^\dag$}} & ItemPosition & 0.6972 & 0.6416 & 0.6369 & 0.7253 & 0.6526 & 0.6115 & 0.7372 & 0.6563 & 0.6013 & 0.7338 & 0.6353 & 0.6105 \\
& Backbone & 0.6900 & 0.6303 & 0.6345 & 0.7156 & 0.6501 & 0.6176 & 0.7415 & 0.6525 & 0.5977 & 0.7399 & 0.6676 & 0.5995 \\
& SASRec & 0.6936 & 0.6232 & 0.6321 & 0.7161 & 0.6508 & 0.6244 & 0.7376 & 0.6590 & 0.6012 & 0.7369 & 0.6400 & 0.6071\\
& DIEN & 0.6877 & 0.6415 & 0.6379 & 0.6986 & 0.6226 & 0.6320 & \underline{0.7419} & 0.6749 & \underline{0.5974} & 0.7408 & 0.6642 & 0.5990 \\
\hline
\textbf{SegRec} & ID & \textbf{0.7276}$^{*}$ & \textbf{0.6713}$^{**}$ & \textbf{0.6092}$^{*}$ & \textbf{0.7339}$^{*}$ & \textbf{0.6723}$^{*}$ & \textbf{0.6020} & \textbf{0.7445} & \textbf{0.6874}$^{**}$ & \textbf{0.5953} & \textbf{0.7462} & \textbf{0.6800} & \textbf{0.5938} \\
\hline
\multicolumn{14}{l}{\parbox{17cm}{$^\dag$ means baselines in video-skip prediction~(Task 1). 
SASRec is the best baselines of GeneralRec in Task 1. AdaGIN and DIEN are the best baselines of ContextRec  for SegMM and KuaiRand datasets in Task 1, respectively. We choose the FREEDOM as baseline for SegMM dataset with visual features.}}
\end{tabular}
\label{tab:task2_results}
\end{table*}

While segment interest modeling is key to video-skip prediction, it also enhances video-level recommendations by providing deeper insights into user preferences. 
We conduct video-level recommendations to evaluate the usefulness of segment interest modeling. Therefore, segment interest scores are introduced intuitively  to existing recommendation models. We propose SegRec, a framework that integrates segment interest for video-level recommendations, as illustrated in Figure~\ref{fig:task2_framework}. 
% 对每一个包含历史浏览的user和target video，经过Sec~\ref{sec:model}, 我们得到user segment interest scores $p_i$. 这代表在不同片段对于是否推荐该video的重要程度。
For each user and target video pair, we compute segment interest scores $p_i$ (as described in Section~\ref{sec:method}), representing the importance of each segment. 
% 在SegRec中，相同的user and video features输入backbone推荐模型，对于每一个segment $i$，推荐模型给出其预测分数$y_i$, 在video-level aggregation中进行加权，得到最终的预测分数:
These scores, along with user and video features, are input into the backbone recommendation model, which predicts segment scores $y_i$. 
The final prediction score
$\hat{y} = \sum_{i=1}^{L} p_{i}^{'} y_{i}^{'} $ is obtained by weighting and aggregating these segment scores, where $\vec{p}^{'} = \mathrm{Softmax}(\vec{p})$. 
Following standard practice, Binary cross-entropy loss is applied with the predicted score $\hat{y}$ and the label $y$, optimizing the backbone model. 

\vspace{-1em}
\subsection{Task Settings}

\subsubsection{Backbones and Baseline Settings}

% As context information is the important input,
We integrated SegRec on several state-of-the-art context-aware methods as backbones, including interaction-based and sequential ones.
The context-aware recommenders include WideDeep~\cite{cheng2016wide} and AdaGIN~\cite{sang2024adagin}, which are from the classic to the latest. 
By adding history information, context-sequential-aware recommenders include DIN~\cite{zhou2018deep} and CAN~\cite{bian2022can}. 
 
As for baseline setting, we use self-information recommend without segments~(Video), segment-level integration by sum~(SegSum) and by learnable weight~(SegAdjust). 
Besides, we serve segment interest scores from task~1, including ItemPosition, the baseline model that is the same as the backbone,  and the best-performing baseline from each category as representatives.

\subsubsection{Implementation Details}

% label 划分
Following previous work~\cite{zhao2023disentangled,li2024modeling,gong2022real}, we define video-level labels for the Click-Through Rate~(CTR) prediction task with ``effective-view'' as follows: whether watch time exceeds the threshold of the corresponding duration bucket. 
In practice, we divide the video duration into 10 buckets and use the median of the view ratio within each bucket as the label threshold.
% data split
To prevent information leakage, the data split strategy follows the same approach as in Task~1. 
% 值得说明的是，所有baseline引入video duration和多模态特征作为视频侧特征，其中多模态特征映射到低维空间。由于多模态的直接引入并不总是能对CTR任务带来效果，我们对引入的特征组合作为input，选取最好的表现，以保证使用信息的公平性。
Notably, all baselines incorporate video duration and multimodal features as video-side features, with multi-modal features mapped to a lower-dimensional space. Since directly introducing multi-modal features does not always improve performance, we select the best-performing combination of features to ensure fairness. 
% 模型调参范围在baseline和我们方法之中也一致，
The model hyperparameter tuning ranges are consistent across both the baseline and our SegRec.
% Detailed settings are available\textsuperscript{\ref{foot:link}}.

\subsection{Video Recommendation Performance} 
\label{sec:Task2results}

Table~\ref{tab:task2_results} shows the recommendation results on SegMM and KuaiRand. 
% 我们的方法在多种backbone上都相较baseline有提升，并且在WideDeep和DIN上都是显著提升。
% 对比Self-information，我们发现直接video recommend的结果已经非常不错，是很多组的best baseline。SegSum和SegAdjust并没有引入更多的知识和信息，很多时候和直接video recommend的表现差不多。但不管如何，引入我们的segments interest scores 对于segAdjust 和SegSum总有显著提升，说明我们得到的segments interest scores的有效性
% Task1-output 类别可以发现，并不是所有segment interests都可以作为权重对CTR任务有贡献，多数baseline反而更差，可以认为其中对于CTR prediction来说倾向于是噪声，突出segments interest modeling的价值。
The experiments demonstrate that our method outperforms the baseline models across different backbones, with significant gains with WideDeep and DIN as backbones. 
% Focusing on the Self-information, direct video recommendation consistently performs well, often outperforming other baselines. SegSum and SegAdjust do not introduce additional knowledge or information, which limits their ability to improve upon the direct video recommendation, resulting in performance that is often comparable. 
Direct video recommendation performs well, as does SegAdjust, which benefits from trained weights for each segment score. In comparison, our approach leverages segment interest scores, which serve as valuable weights for individual segments, leading to targeted improvements in overall model performance. The consistent enhancement underscores the utility and effectiveness of segment interest modeling.

When applying the output of Task~1's baseline as the weight, we observed that their performance generally aligns with its performance in Task~1. For example, AdaGIN, as the best baseline for SegMM in Task~1, is also the best baseline when used for video recommendations. 
However, not all segment interests contribute positively to the video recommendation task, suggesting that some segment interests may introduce noise into recommendation. In contrast, our segment-level interest modeling always outperforms, further demonstrating the effectiveness of accurate interest scores.
Regarding modality fusion, combining modalities~(both) yields the best results, with ID consistently outperforming visual modality. 

To be noted, our framework is designed primarily to validate the effectiveness of segment interest scores. We find the utility of these scores in improving performance while maintaining a directed and explainable approach. This provides insight into how segment-level data can be served effectively and interpretable without overcomplicating the recommendation process. 

% \subsection{Ablation Study} 
% \label{sec:Task2ablation}

% 与Task1 相同，我们探究不同模态对结果的影响。Table~\ref{tab:task2_SegMM_results} 也展示了不同模态的segments interest scores的推荐效果。总体而言，both总之效果最好的，ID和visual表现差别不大。说明fuse不同模态能对下游的推荐任务带来更好的表现。
% Similar to Task 1, we investigate the impact of different modalities on the results. Table~\ref{tab:task2_results} shows the recommendation performance of segment interest scores for different modalities. 
% Overall, the combined modality~(both) performs the best, with ID and visual modalities showing similar results.
% This indicates that fusing different modalities can improve performance in downstream recommendation tasks.

%% file: Main/6.Case.tex
% And Limitation
\vspace{-1em}
\section{Case Study and Discussions}
\label{sec:case}

\begin{figure}[t]
    \centering
    \setlength{\abovecaptionskip}{0.2cm}
    \setlength{\belowcaptionskip}{-0.5cm}
    \includegraphics[width=8.6cm]{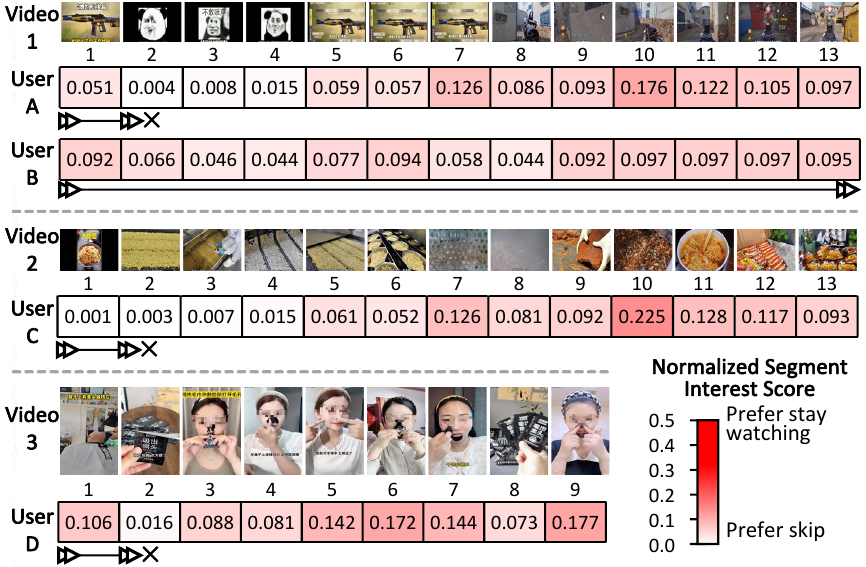}
    \caption{Cases with represented images, position indices, our model predicted interest scores~(normalized), and user behaviors~(unknown to the model). $\rhd$ \kern-0.7em{\colorbox{white}{$\rhd$}}\kern-0.3em{-}\kern-0.15em{-}\kern-0.15em{-} \kern-0.35em$\rhd$ \kern-0.7em{\colorbox{white}{$\rhd$}} represents keep viewing and \twemoji{multiply} is video skipping.} % triangle play, true label, predict CTR(task2x), (a) (b) ---, V1UA Video1UserA  Video recommendation scores with AdaGIN as backbone are shown for every case~(e.g., Case~1A represents Video~1 for User~A). ``baseline'' do not use of segment interest scores. 
    \label{fig:case} 
\vspace{-1em}
\end{figure}

% To eveluate the effectiveness of segment-level interest modeling, we show several cases to visually present the model's output, highlighting the model's predictions of segment interest based on segment images, position indices, and user behaviors.
Previous sections proposed segment-level interest modeling and evaluated it on two tasks with improved accuracy. As shown in Figure~\ref{fig:case}, several cases illustrate the model's performance by highlighting predicted segment interest based on segment images, position indices, and user history. 
% 对于每个case: （1）segment代表图像及用户对于这些video的兴趣预测对应图像位置排列（强调是模型未知用户行为的预测结果）。（2）用户真实行为，包括用户播放到下滑。（3）兴趣分数归一化（Proportional Normalization）和 CTR任务中的模型预测分数。
Each test case presents segment images in sequence. %, with predicted user interest corresponding to the segment positions. 
The predicted interest scores, obtained through proportional normalization of model outputs, are visualized as a heatmap, with user viewing and skipping behaviors shown below. 
% The higher the prediction score (with the darker color), the more likely the user is to stay at this video and vice versa. 
%The predictions are based on model outputs, with user behaviors unknown to the model. 
% with user viewing and skipping behaviors are displayed below the scores.  
% Below the predicted interest scores, actual user behaviors are displayed, including the point where users skip to the next video. The interest scores are normalized using proportional normalization, and CTR predictions are included. 
% The CTR prediction scores in video recommendation for this cases are also given, showing considerable improvements~(closer to the CTR label) of our methods in most cases. 

% Case1&2 和 case3&4 都是同一个视频在不同用户上的表现，对于Case1&2的video，训练集的12个交互全为用户在第一个片段就滑走，user1也同样，并且兴趣最低的片段即为用户离开片段。user2则是在滑走前的四个分数持续较低，长期低兴趣的累积使得该用户在第五个片段滑走。 Case3&4是两个用户对一个介绍游戏中枪的视频的测试样例。模型预测兴趣显式，两人均在第2-4段表情包持续存在时兴趣较低。用户3在表情包片段的兴趣尤其低，对应在第二个片段滑走，尽管用户4的兴趣也有变化，但相对均匀，对应完播行为
% Case~1A and 1B represent the same video across different users. 
Case~1A~(short for user~A interacting with video~1, similarly for other cases) and 1B demonstrate the model’s user-aware generalization ability, with two different prediction-behavior pairs for video~1.  
The video spends a long period~(segments~2-4) on emoticons without substantive information, which distracts most users. The model captures this effectively, assigning low interest scores for both users. 
% Comparing two users, our model outputs even lower interest scores at the emoticon segments for user~A than user~B based on their personalized history views, which shows remarkable consistency with their actual behaviors: skipping at segment~2 and watching the whole video, respectively. 
Based on personalized histories, the model outputs even lower scores for user~A than user~B, aligning with actual behaviors: user~A skips at segment~2, while user~B watches the full video.  
We also examine the video recommendation prediction scores in Task~2 for these two cases using AdaGIN as the backbone. 
\textbf{SegRec} improves recommendation scores~(closer to true labels) by incorporating segment interest: from 0.2144 to 0.1889 for case~1A~(label=0), and from 0.3601 to 0.5936 for case~1B~(label=1), compared to the baseline without interest scores. This demonstrates the value of segment interest scoring.
% 对于User B的预测相对平缓，turns out user B完播
% However, scores user~B maintains steady interest and completes the video. 
% In case~1\&2, all 12 interactions for this video in training set are skipped at the 1st segment, with user~1 also skipping here with the lowest interest. User~2, however, shows continuously low interest over the first four segments and eventually skips at the 5th. 
% Video~1 is about introducing a gun in a game. The model predicts low interest during the emoticon segments, especially for user~A, who skips at the 2nd segment, while user~B maintains steady interest and completes the video. Cases show that our predicted scores align with the actual user behaviors.

% \setlength{\fboxsep}{0pt}

% case5和case6均为对于模型的cold item。Case5是关于方便面的介绍，存在场景切换——视频前半部分是加工，后半部分是食品烹饪及美食展示，模型预测用户3对视频前面的兴趣较低，而对后面的兴趣较高，美食是她的兴趣但她却早已滑走，对应兴趣segment level预测的价值：帮助系统hightlight给用户这个视频中用户会喜欢的部分，而不会因为所处位置靠后而错过。
% Case6是一个广告视频，模型预测出用户在第2和8个片段兴趣最低，对应该片段呈现出广告图片，用户在出现广告时滑走，体现出模型对于多模态信息的捕捉能力。
Case~2C and 3D involve cold items, showcasing the model's adaptability and understanding of multimodal information. 
Case~2C is an example of scene transition: segments 1–9 show the instant noodle production process, while later segments focus on cooking.  
The model assigns higher scores to the latter, inferring the user’s preference for food content. 
However, user~C skipped early at segment~2, missing possible interest parts, indicating segment-level interest modeling could help recommendation by parts of the videos instead of entire ones. 
% Video~2 is about instant noodles, where a scene switch occurs: the beginning focuses on production, and the rest on cooking and food display. The model predicts user~C shows low interest in the beginning and high interest in the latter. User~C skips early, missing the part he would have liked, demonstrating how segment-level interest prediction can help recommend the most preferred part of a video, even if it appears later. 
In advertisement video~3, segments~2 and 8 show product's appearance, while others introduce its usage. 
The model outputs a relatively low interest score at the product segments, meeting the user behavior of skipping at segment~2. 
% Video~3 features ads, where the model predicts low interest in segments~2 and 8, which display products's appearance. The user skips these segments, demonstrating the model's ability to capture multimodal information and predict user behavior accurately. 

% A comparison of the video recommendation task with CTR labels for the cases is also reported in Figure~\ref{fig:case}. 
% SegRec, with integrating segment-level interest scores, triumphs over baseline in most cases with even a correction of false predictiotn (case~1B), showing the effectiveness of segment interest modeling. 

% Our model shows its ability to capture personalization and multimodal information. 
% Cases showcase an optimistic consistency between segment-level modeling and user behavior, containing great potential for complex tasks beyond traditional recommendation,
% It also highlights its potential in video personalized thumbnail generation and video editing suggestions. % 很需要说清楚这部分价值：
% （1）个性化缩略视频是在首页呈现的，吸引用户点击进视频。
% （2）个性化剪辑是将用户兴趣低的部分自动快进视频，或剪辑出用户感兴趣的内容
The results highlight the potential of our model in \textbf{video personalized thumbnail generation},  allowing for the display of interest thumbnails on users' homepages to attract clicks. 
It enables \textbf{personalized video editing suggestions}, where uninterested sections are automatically fast-forwarded, and content is tailored to users' specific interests. 
Unfortunately, those untapped tasks were not included in this work due to the lack of valid datasets. We sincerely call upon more informative datasets for future studies. 
% 提供的新的价值：我们不仅推荐用户喜欢的内容，而且针对用户的个性化兴趣提供内容
% Beyond recommending content based on users' general preferences, our work introduces a new perspective by delivering content tailored to users' personalized and dynamic interests. 

% As we describe in

% % 展示case
% Case

% % 广告视频
% 1. advertisement

% % 剧情视频
% 2. story/scene

% % 和群体不一致的popular video
% 3. personalized case in popular video 

% Emphasize potential applications

% reveal a light

%% file: Main/7.Conclusion.tex
% \section{Conclusions and Future Works}
\section{Conclusions}
\label{sec:conclusion}

To our knowledge, this is the first research in the recommendation field focusing on segment-level dynamic user interest. 
We propose a novel modeling method to extend the paradigm of the recommendation system from the video level to the segment level. 
Consisting of hybrid representation, multi-modal encoder, and segment interest decoder, the model excelled at video skip prediction and recommendation tasks, evaluated by experiments across two datasets and demonstrated by several cases, showing its ability in personalized recommendation and potential for more complicated tasks. 
% Unlike existing methods that focus on video-level features, we propose a novel approach that captures dynamic user interests at the segment level. 
% By leveraging a hybrid representation and a segment interest decoder, our model improves both video-skip prediction and recommendation accuracy by considering temporal shifts in user engagement. 
% Experiments on two tasks~(video-skip prediction and video recommendation) across two datasets demonstrate the effectiveness of our segment interest modeling, highlighting its ability to enhance personalized recommendations. 

We hope this work can pave the way beyond traditional content recommendation by focusing on delivering content tailored to users' evolving interests. 
%The dataset we released is believed to provide a valuable resource for future research. 
With the development of valid datasets, we aim to conduct more research to explore the field of segment characterization, bridging the gap between human cognition and innovative, practical video applications in the future. 
% Our work paves the way to personalized video experiences by enabling dynamic thumbnail generation for user homepages and offering personalized video editing suggestions, such as fast-forwarding uninterested sections. 
% It goes beyond traditional content recommendations by focusing on delivering content tailored to users' evolving interests. The open-sourcing of our dataset can provide a valuable resource for future research. Overall, segment-level interest modeling holds great potential to enhance user experience and recommendation systems on short video platforms. We hope to inspire more researchers to explore this field, bringing deeper insights and practical applications. 